\RequirePackage{fix-cm}
\documentclass[twocolumn,epjc3]{svjour3}  
\smartqed  
\RequirePackage{graphicx}
\usepackage{amsmath} 
\usepackage{amssymb}
\usepackage{cancel}
\usepackage[normalem]{ulem}
\usepackage{xcolor}
\journalname{Eur. Phys. J. C}
\begin{document}

\title{Temporary EHBL-like behavior of Markarian 501 during July 2014 VHE flaring
}



\author{Sarira Sahu\thanksref{e1,addr1}
        \and
        A. U. Puga Oliveros\thanksref{e2,addr1}
        \and
        D. I. Páez-Sánchez\thanksref{e3,addr1}
        \and
        G. Sánchez-Colón\thanksref{e4,addr2}
        \and
        Subhash Rajpoot\thanksref{e5,addr3}
        \and
        M. E. Iglesias Martínez\thanksref{e6,addr4}
        \and
        Jose Guerra Carmenate\thanksref{e7,addr4}
        \and
        P. Fernández de Córdoba\thanksref{e8,addr4}
        \and
        Gaetano Lambiase\thanksref{e9,addr5,addr6}
}

\authorrunning{S. Sahu et al.} 

\thankstext{e1}{e-mail: sarira@nucleares.unam.mx}
\thankstext{e2}{e-mail: angel.puga@correo.nucleares.unam.mx}
\thankstext{e3}{e-mail: diana.paez@correo.nucleares.unam.mx}
\thankstext{e4}{e-mail: gabriel.sanchez@cinvestav.mx}
\thankstext{e5}{e-mail: Subhash.Rajpoot@csulb.edu}
\thankstext{e6}{e-mail: miigmar@upv.es}
\thankstext{e7}{e-mail: jguecar@doctor.upv.es}
\thankstext{e8}{e-mail: pfernandez@mat.upv.es}
\thankstext{e9}{e-mail: lambiase@sa.infn.it}

\institute{Instituto de Ciencias Nucleares, Universidad Nacional Autónoma de México, Circuito Exterior S/N, C.U., A.P. 70-543, CDMX 04510, México \label{addr1}
           \and
           Departamento de Física Aplicada, Centro de Investigación y de Estudios Avanzados del IPN, Unidad Mérida. A.P. 73, Cordemex, Mérida, Yucatán 97310, México \label{addr2}
           \and
           Department of Physics and Astronomy, California State University, 1250 Bellflower Boulevard, Long Beach, CA 90840, USA \label{addr3}
           \and
           Instituto Universitario de Matemática Pura y Aplicada, Universitat Politècnica de València, Camino de Vera s/n, 46022 Valencia, Spain \label{addr4}
           \and
           Dipartimento di Fisica “E.R. Caianiello”, Università di Salerno, Via Giovanni Paolo II, I-84084 Fisciano (SA), Italy \label{addr5}
           \and
           INFN, Gruppo collegato di Salerno, Via Giovanni Paolo II, I-84084 Fisciano (SA), Italy \label{addr6}
}

\date{Received: date / Accepted: date}

\maketitle

\begin{abstract}
Markarian 501, a BL Lac object well-known as a high energy gamma-ray source, has exhibited several epochs of very high energy (VHE) gamma-ray flaring events when its synchrotron peak frequency shifted above $10^{17}$ Hz, a signature of extreme behavior. From July 16 to July 31, 2014 such flaring events were observed for 15 days by various telescopes. On July 19 (MJD 56857.98), the X-ray outburst from the source was at its highest and on the same day an intriguing narrow peak-like feature around 3 TeV was observed by the MAGIC telescopes, a feature inconsistent with standard interpretations. Using the well-known two-zone photohadronic model, we study these VHE gamma-ray spectra on a day-by-day basis and offer explanation.
Our two-zone photohadronic scenario shows that, on MJD 56857.98, the peak-like feature appears at a cutoff energy of $E^c_{\gamma}=3.18$ TeV. Below this energy the VHE spectrum increases slowly and is in high emission state. However, for $E^c_{\gamma}\, > 3.18$ TeV, the spectrum falls faster, resulting in a mild peak-like feature, not prominent enough as claimed by the MAGIC collaboration.



\end{abstract}


\section{Introduction}

Markarian 501 (Mrk 501) is a high-frequency peaked BL Lac object (HBL) at a redshift of $z=0.034$ ~\cite{1975ApJ...198..261U}. It is the second well-known bright HBL after Markarian 421 (Mrk 421) and is one of the brightest extragalactic sources in the X-ray/TeV sky. After its discovery in very high energy (VHE) ($> 100$ GeV) by the Whipple telescopes in 1996 ~\cite{1996ApJ...456L..83Q}, several major outbursts in multi-TeV have been observed and have been the target of many multiwavelength campaigns mainly covering VHE flaring activities ~\cite{2011ApJ...727..129A,2015A&A...573A..50A,2015ApJ...812...65F,2017A&A...603A..31A,2018A&A...620A.181A}. Mrk 501 is one of the few blazars which can be monitored even during the low emission periods. To understand different emission mechanisms and their characterizations, it is essential to have long-term and multiwavelength observations of the spectral energy distribution (SED) of the blazars. Also, simultaneous correlation studies in different wavelengths of the SED during the flaring can help us to constrain the emission mechanisms. Motivated by these goals, several multiwavelength and multi-instrument observation campaigns were organized.
In 1997, from March to October, Mrk 501 had an unprecedented flare in VHE gamma\textcolor{magenta}{-}rays and the integral flux reached up to four times the magnitude of the Crab Nebula (CU) flux \cite{2000ApJ...536..742P}. Also, it was extremely active in the X-ray region during this period and the synchrotron peak had a dramatic shift to energies above 100 keV, which is more than two orders of magnitude compared to that of the typical non-flaring state.

During May to July 2005, Major Atmospheric Gamma-Ray Imaging Cherenkov (MAGIC) telescopes made observations for 30 nights, and flaring above 100 GeV was observed with an order of magnitude flux variation \cite{2007ApJ...669..862A}. On the nights of June 30 and July 9, flux-doubling in the VHE band was observed within a time span of about 2 minutes. This is the fastest flux variation ever observed from Mrk 501 since its discovery. During this observation period, the synchrotron peak and the VHE peak were shifted towards higher energy limits. As a part of the long term multiwavelength campaign, Mrk 501 was observed by $\sim 30$ different instruments covering nearly the entire electromagnetic spectrum \cite{2017A&A...603A..31A,2016A&A...594A..76A} from March 15 to August 1, 2009. 

Between March and July of 2012, another multiwavelength campaign was undertaken when more than 25 different telescopes, including MAGIC, and Very Energetic Radiation Imaging Telescope Array System (VERITAS) took part and observations at energies $>\,0.2$ TeV were made~\cite{2018A&A...620A.181A}. The highest peak flux was recorded on June 9, and most importantly, the X-ray spectrum and the VHE gamma-ray spectrum were observed to be extremely hard. During this period, it is observed that the synchrotron and the Synchrotron Self-Compton (SSC) peaks were above 5 keV and above 0.5 TeV respectively. The TeV Atmospheric Cherenkov Telescope with Imaging Camera (TACTIC) at Mt Abu, Rajastan, India, observed Mrk 501 from 15 April to 30 May, 2012 at energies above 850 GeV for 76.0 h. Between 22 to 27 May, 2012, TACTIC telescope observed relatively high VHE flux in the energy range between 850 GeV and 17.24 TeV~\cite{2017NewA...54...42C}.

Mrk 501 was monitored in the year 2014 with the First G-APD Cherenkov Telescope (FACT) \cite{2013JInst...8P6008A,2014JInst...9P0012B,2015arXiv150202582D}. Several high flaring states were observed. This triggered the High Energy Stereoscopic System (HESS) to undertake further observations. Enhanced VHE gamma-ray flux, comparable to the exceptional flare of 1997, was observed on the night of June 23-24, 2014 (MJD 56832) with the HESS phase-II telescopes. The energy of the VHE photons extended up to 20 TeV and the flux had a rapid variability. Estimates indicate that this is the highest flux ever observed from Mrk 501 with the HESS telescopes \cite{2019ApJ...870...93A}. 

Again, in July of 2014, flaring in X-ray and VHE gamma-ray were observed for about two weeks (from 16th to 31st)~\cite{2020A&A...637A..86M}. The X-ray flaring was found to be exceptionally high, similar to the historic 1997 flaring. This amounts to the largest flux ever detected so far with the X-ray Telescope (XRT) on board the \textit{Neil Gehrels Swift} Observatory \cite{2004ApJ...611.1005G} in its operational history. Also, the VHE gamma-ray emission above 0.15 TeV during this time was high, about 0.5 to 2 times of the CU flux. On July 19 (MJD 56857.98), the day of the highest X-ray activity, a narrow peak-like feature at $\sim 3$ TeV was observed by MAGIC telescopes \cite{2020A&A...637A..86M}.

In the Universe, the most dominant sources of extragalactic gamma rays are the blazars~\cite{2017SSRv..207....5R}. Their SED is characterized by two non-thermal peaks~\cite{2010ApJ...716...30A}, and the most popular models interpret these peaks as the synchrotron peak is in the infrared to X-ray frequency range and the SSC peak is in the X-ray to gamma-ray range \cite{1998MNRAS.299..433F,1998MNRAS.301..451G,1999APh....11..159U}. The BL Lac objects belong to the blazar family and are classified according to the position of their synchrotron peak frequency $\nu^{p}_s$ in the SED;  as low-frequency peaked (LBLs, $\nu^{p}_s < 10^{14}\, \mathrm{Hz}$), intermediate-frequency peaked (IBLs, $10^{14}\, \mathrm{Hz} < \nu^{p}_s < 10^{15}\, \mathrm{Hz}$), high-frequency peaked (HBLs, $10^{15}\, \mathrm{Hz} < \nu^{p}_s < 10^{17}\, \mathrm{Hz}$~\cite{2010ApJ...716...30A,1995ApJ...444..567P,2013ApJ...768...54B}) and extreme high-frequency peaked (EHBLs, $\nu^{p}_s > 10^{17}\, \mathrm{Hz}$~\cite{2001A&A...371..512C}) with extreme spectral properties.

Low luminosity and limited variability of EHBLs hinder the detection of their gamma-rays by the current generation Imaging Atmospheric Cerenkov Telescopes (IACTs). Thus, only a handful of EHBLs have been detected so far and well-known among them are 1ES 0229+200, 1ES 0347-232, RGB J0710+591
and 1ES 1101-232~\cite{2018MNRAS.477.4257C}. It is observed that at least three nearby extensively studied
HBLs, Mrk 421, Mrk 501 and 1ES 1959+650 have shown temporary EHBL (tEHBL) like behavior during several flaring epochs~\cite{2020ApJ...901..132S,2021ApJ...906...91S,2021ApJ...914..120S}. During tEHBL-like behavior, the synchrotron peaks have shifted above $10^{17}$ Hz and the TeV spectra are harder. Also, due to the complex behavior of these sources with different spectral properties at VHE gamma rays, the possibility of different subclasses within the EHBL class cannot be ruled out~\cite{2019MNRAS.486.1741F}.


With the one-zone leptonic SSC model, it is difficult to explain the VHE spectrum of EHBL, since the leptonic model predicts a softer SSC spectrum in the Klein-Nishina regime, and is contrary to observations~\cite{2009A&A...504..821P}. One can circumvent this problem by adopting 
very large values of electron Lorentz factor, bulk Lorentz factor and a very low magnetic field~\cite{2018A&A...620A.181A,2020A&A...638A..14M}. Several alternative models such as the two-zone leptonic model, the inverse Compton scattering of relativistic electrons with the cosmic microwave background, the spine-layer structured jet model, and different hadronic models are proposed to explain these spectral behavior~\cite{2018A&A...620A.181A,2008ApJ...679L...9B,2019MNRAS.490.2284M}.



VHE gamma rays from the extragalactic sources get attenuated by interacting with the extragalactic background light (EBL) and this attenuation is energy dependent. The shape of the observed spectrum changes at very high energies, so, EBL corrections to the observed spectrum have to be taken care of~\cite{2008A&A...487..837F,2011MNRAS.410.2556D,2014MNRAS.438.3255T,2015arXiv151205080T}. EHBLs have extreme spectral properties in TeV energy range, and probably these are ideal probes for EBL.

Mrk 501 had VHE flaring during 2005 and 2012 when the synchrotron peak had shifted temporarily above 1 keV and the SSC peak also shifted to higher energy, a clear signature of EHBL-like behavior. The shift of the synchrotron peak and the SSC peak to higher energies in the
multiwavelength SED generally translate to a particularly hard
X-ray and a very-high-energy gamma-ray spectra, difficult to explain using leptonic model.

Different temporal behaviour has been observed among the EHBLs. While the EHBLs, 1ES 0229+200 and 1ES 034-232 exhibit extreme properties constantly, the well-known HBLs, such as Mrk 421, Mrk 501 and 1ES 1959+650 enter to the EHBL family as temporary members, with peculiar spectral properties~\cite{2018A&A...620A.181A,2007ApJ...669..862A,2020A&A...638A..14M,2015A&A...578A..22A}. The VHE spectrum of a tEHBL falls faster in the high energy regime (above a cutoff energy) as compared to low energy regime.
This behavior is not observed in HBLs and EHBLs and can not be explained using the conventional photohadronic model. Thus, the photohadronic model was extended to two-zones, which could explain the VHE spectra extremely well~\cite{2020ApJ...901..132S,2021ApJ...906...91S,2021ApJ...914..120S}. The VHE flaring event of July 2014 of Mrk 501 was very similar to the above flaring events. So, in this work, our goal is to explain the VHE flaring events observed during July 16 to July 31, 2014, and particularly, to interpret the unusual peak-like feature at about 3 TeV observed on July 19 (MJD 56857.98) in the context of the two-zone photohadronic model.


The paper is organized as follows. In section 2, the two-zone photohadronic model is briefly discussed. In section 3, we discuss the flaring events of Mrk 501 during July 16 to July 31, 2014 (MJD 56854--56869). A detailed analysis of the VHE events observed by MAGIC for about two weeks is discussed in section 4 and a brief discussion is given in section 5. 


\section{Two-Zone Photohadronic Model}

The synchrotron peak and the SSC peak of a BL Lac object are conventionally explained through
the leptonic models. In these models, the region where the photons are emitted is a blob of comoving radius $R'_b$ (where $'$ implies comoving frame), with a bulk Lorentz factor $\Gamma$ and a Doppler factor $\mathcal{D}$. For HBLs, $\Gamma \simeq \mathcal{D}$.

In the photohadronic scenario, accelerated protons are injected in the jet with a differential spectrum given as $dN/dE_p\propto E^{-\alpha}_p$,
where the proton spectral index $\alpha \ge 2$~\cite{1993ApJ...416..458D}. These protons interact with the background seed photons to produce $\Delta$-resonances, $p\gamma\rightarrow\Delta^+$. These background seed photons are produced from the leptonic processes (in the leptonic models). Decay of the $\Delta^+$ to neutral pion produces observed gamma-rays. The kinematical condition to produce $\Delta^+$ is given by \cite{Sahu:2019lwj,2019ApJ...884L..17S},


\begin{equation}
E_p\, \epsilon_\gamma = \frac{0.32\, \Gamma^2}{(1+z)^{2}}~\mathrm{GeV}^2,
\label{eq:kinproton}
\end{equation}
where $E_p$ and $\epsilon_{\gamma}$ are respectively the proton energy and the seed photon energy in the observer's reference frame. In Eq.(\ref{eq:kinproton}), $z$ is the redshift of the source. The energy of the observed gamma-ray $E_{\gamma}$ from the $\pi^0$ decay carries 10\% of the original proton energy.

It is observed that in the standard jet scenario of the BL Lac objects, the $p\gamma\rightarrow\Delta^+$ process is inefficient due to the low background photon density~\cite{2017MNRAS.465.3506P}.To avoid this problem and to explain the VHE gamma-ray events from HBLs, a double jet structure is assumed in the context of the photohadronic model~\cite{Sahu:2019lwj}. In this model, it is assumed that during the VHE flaring of the source, an internal compact jet is formed within the larger jet~\cite{2016EPJC...76..127S,2018EPJC...78..484S}. The inner jet has radius $R'_f$ and the outer jet has radius $R'_b$ and $R'_f < R'_b$. Also, the photon density in the inner jet region, $n'_{\gamma,f}$, is larger than the photon density in the outer jet region, $n'_{\gamma}$ ($n'_{\gamma,f}\, \gg \, n'_{\gamma}$). During the VHE flaring, the accelerated high energy protons interact with the seed photons in the inner jet region to produce $\Delta$-resonance. Due to high density photons in the inner jet region, the efficiency of the production of $\Delta$-resonance is relatively high as compared to the standard jet scenario and this avoids the necessity for super-Eddington luminosity of protons.

Since $R'_f < R'_b$, the photon density in the inner jet region $n'_{\gamma,f}$, can not be observed directly. But, in the outer region the photon density $n'_{\gamma}$, can be calculated from the observed flux.
Also, due to the adiabatic expansion of the inner jet, its photon density decreases when it crosses into the outer jet. To connect the photon densities in the inner and outer regions, we assume a scaling relation given by \cite{2019ApJ...884L..17S,2013PhRvD..87j3015S}
\begin{equation}
\frac{n'_{\gamma,f}(\epsilon_{\gamma,1})}{n'_{\gamma,f}(\epsilon_{\gamma,2})} 
\simeq 
\frac{n'_{\gamma}(\epsilon_{\gamma,1})}{n'_{\gamma}(\epsilon_{\gamma,2})}.
\label{eq:scaling}
\end{equation}
The above relation shows that during the multi-TeV gamma-ray flaring in a blazar, the ratios of the photon densities at energies $\epsilon_{\gamma,1}$ and $\epsilon_{\gamma,2}$ in the inner jet and the outer jet regions are almost the same. Thus, by using the relation in Eq. (\ref{eq:scaling}), we can express the unknown photon density in the inner jet region in terms of the observed photon density in the outer jet region.

From the previous analysis of the VHE spectra of dozens of HBLs in the photohadronic model, we found that to produce the observed $E_{\gamma}$, the background seed photon energy $\epsilon_{\gamma}$ always lies in the low energy tail region of the SSC spectrum which satisfies Eq. (\ref{eq:kinproton}), and the SSC flux is a perfect power-law, given by $\Phi_{SSC}\propto \epsilon^{\beta}_{\gamma}\propto E^{-\beta}_{\gamma}$, 
with $\beta$ the photon spectral index \cite{Sahu:2019lwj}. 
However, the low energy tail region of the SSC spectrum is model dependent as there is no observation in this energy band.
The observed GeV-TeV photon flux from a HBL is given as \cite{Sahu:2019lwj,2019ApJ...884L..17S}
%
%

\begin{align}
F_{\gamma,\mathrm{obs}}(E_{\gamma}) 
&= F_0 \left( \frac{E_\gamma}{\mathrm{TeV}} \right)^{-\delta + 3} 
   e^{-\tau_{\gamma\gamma}(E_\gamma, z)} \nonumber\\
&= F_{\gamma,\mathrm{in}}(E_{\gamma})\, 
   e^{-\tau_{\gamma\gamma}(E_\gamma, z)}.
\label{eq:fluxgeneral}
\end{align}

In the above equation, $F_0$ is the flux normalization, $F_{\gamma,in}$ is the intrinsic VHE gamma-ray flux and the exponential factor corresponds to the survival probability of VHE photon due to its interaction with the EBL photons which has the optical depth $\tau_{\gamma\gamma}$. For the EBL correction to the VHE spectrum, here, we use the EBL model of Franceschini \textit{et al} \cite{2008A&A...487..837F}. The exponent $\delta=\alpha+\beta$ and for HBLs, it is in the range $2.5\le \delta \le 3.0$. From detailed analysis of several VHE spectra of HBLs it was observed that the flaring states can be in three categories, viz,
(i) very high emission state with $2.5\le\delta\le 2.6$, (ii) high emission state with $2.6<\delta<3.0$ and (iii) low emission state with $\delta=3.0$ \cite{2019ApJ...884L..17S}. But for a tEHBL spectrum, it was shown that the single power-law behavior of the form $\Phi_{SSC}\propto E^{-\beta}_{\gamma}$ does not work\cite{2020ApJ...901..132S,2021ApJ...906...91S,2021ApJ...914..120S}. Thus, two power-laws for the background seed photons in the low energy tail region is assumed in the form~\cite{2020ApJ...901..132S},

\begin{equation}
\Phi_{SSC}\propto
 \left\{ 
\begin{array}{cr}
E^{-\beta_1}_{\gamma}
, & \quad 
 \mathrm{100\, GeV\,} \lesssim E_{\gamma} \lesssim E^{c}_{\gamma}
\\ E^{-\beta_2}_{\gamma} ,
& \quad   E_{\gamma}\gtrsim E^{c}_{\gamma}
\\
\end{array} \right.,
\label{eq:sscflux}
\end{equation}

with the spectral indices $\beta_1 \neq\beta_2$ ($\beta_1,\beta_2 > 0$) and $E^{c}_{\gamma}$ is the cutoff energy scale around which the transition between zone-1 and zone-2 takes place. The value of $E^{c}_{\gamma}$ is different for different spectra. Thus, using Eq.(\ref{eq:sscflux}), the observed spectrum for a tEHBL can be expressed as
\begin{equation}
F_{\gamma, obs}=
e^{-\tau_{\gamma\gamma}}
\begin{cases}
 F_1 \left ( \frac{E_{\gamma}}{TeV} \right )^{-\delta_1+3}, & \,(\text{zone-1}) \\ 
F_2 \left ( \frac{E_{\gamma}}{TeV} \right )^{-\delta_2+3},
&  \,\,  (\text{zone-2})
 \end{cases},
\label{eq:flux}
\end{equation}
were $F_1$ and $F_2$ are the normalization constants and $\delta_i=\alpha+\beta_i$ ($i=1,2$) are the free parameters in the photohadronic model that are to be adjusted by fitting to the observed VHE spectrum. In Eq.(\ref{eq:flux}), the zone-1 corresponds to the energy range  $\mathrm{100\, GeV\, \lesssim E_{\gamma} \lesssim E^{c}_{\gamma}}$ and the zone-2 corresponds to 
$\mathrm{E_{\gamma}\gtrsim E^{c}_{\gamma}}$.
Previously, the study of tEHBL-like behavior of Mrk 421, Mrk 501 and 1ES 1959+650 in the context of two-zone photohadronic model has shown $2.5\le \delta_1 \le 3.0$ and $3.0 < \delta_2 \le 3.5$~\cite{2020ApJ...901..132S,2021ApJ...906...91S,2021ApJ...914..120S}.

\section{VHE flaring of Markarian 501 during 2014}

Mrk 501 was observed in multiple wavelengths by various instruments from July 16 to 31, 2014. During this period, the MAGIC telescopes observed for 13.5 h under dark and good atmospheric conditions~\cite{2020A&A...637A..86M}. The X-ray activity observed by the {\it Neil Gehrels Swift} Observatory was at its highest during its operational history and the spectra observed by different X-ray telescopes were very hard, similar to the large historic flare of 1997 when the synchrotron peak was shifted above 100 keV \cite{1998ApJ...492L..17P,2001ApJ...554..725T}, an indication of tEHBL-like behavior. 
It is interesting to note that the flux variation during July 2014 in the energy regime covering the radio, optical and GeV energies was mild, but the enhanced emission in X-ray and VHE gamma-ray was significantly correlated ($>3\sigma$). Also, it was observed that the strength and the significance of the correlation increased for increasing energy in X-rays. For 0.3-2 keV the correlation was $\sim 3\sigma$ and for 2-10 keV the correlation was $\sim 4\sigma$ respectively~\cite{2020A&A...637A..86M}. Temporal evaluation of the fluxes in X-ray and VHE gamma-ray were characterized on a day-by-day basis from July 16 to July 31, 2014 (MJD 56854--56869). Harder-when-brighter trends in X-ray spectra were observed, but the VHE spectra observed by MAGIC did not show such behavior. 

During the X-ray outburst, it was observed that the VHE gamma-ray flux above 0.15 TeV varied from $\sim 0.5$ to $\sim2$ times the CU flux and no intra-night variability was observed in VHE gamma-rays during this two weeks flaring period.
On 2014 July 19 (MJD 56857.98) the highest X-ray flux above 2 keV was recorded by {\it Swift}, while the MAGIC telescopes observed a rare unusual narrow spectral feature at $\sim 3$ TeV in the VHE spectrum. This intriguing peak-like feature is inconsistent at more than $3\sigma$ with the standard analytic functions such as: Power Law, log-parabola and log-parabola with exponential cutoff \cite{2020A&A...637A..86M}, widely used to fit the VHE spectra of blazars. The observed broadband SED for 15 consecutive days, from July 16 to July 31, 2014, were explained using one-zone SSC models. Assuming that the narrow feature at $\sim 3$ TeV to be real, several theoretical scenarios were proposed to explain it~\cite{2020A&A...637A..86M}. Also, this narrow spectral feature on MJD 56857.98 is modeled by using the leptonic and the hadronic models~\cite{2021MNRAS.508.4038H,2024A&A...685A.110P}.


\begin{figure}
  \begin{center} 
        \includegraphics[width=0.46\textwidth]{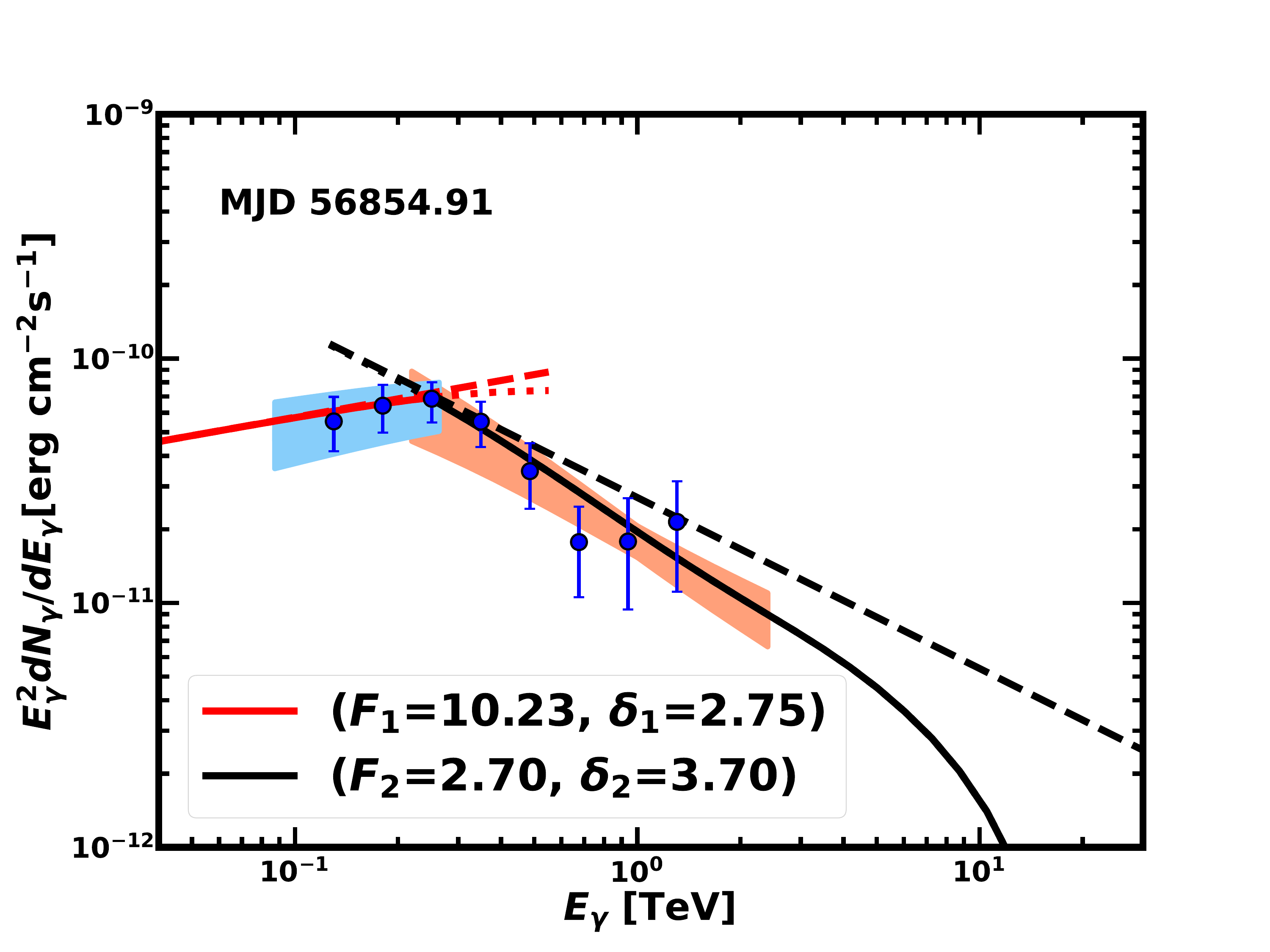}
        \label{Fig1a}
\includegraphics[width=0.46\textwidth]{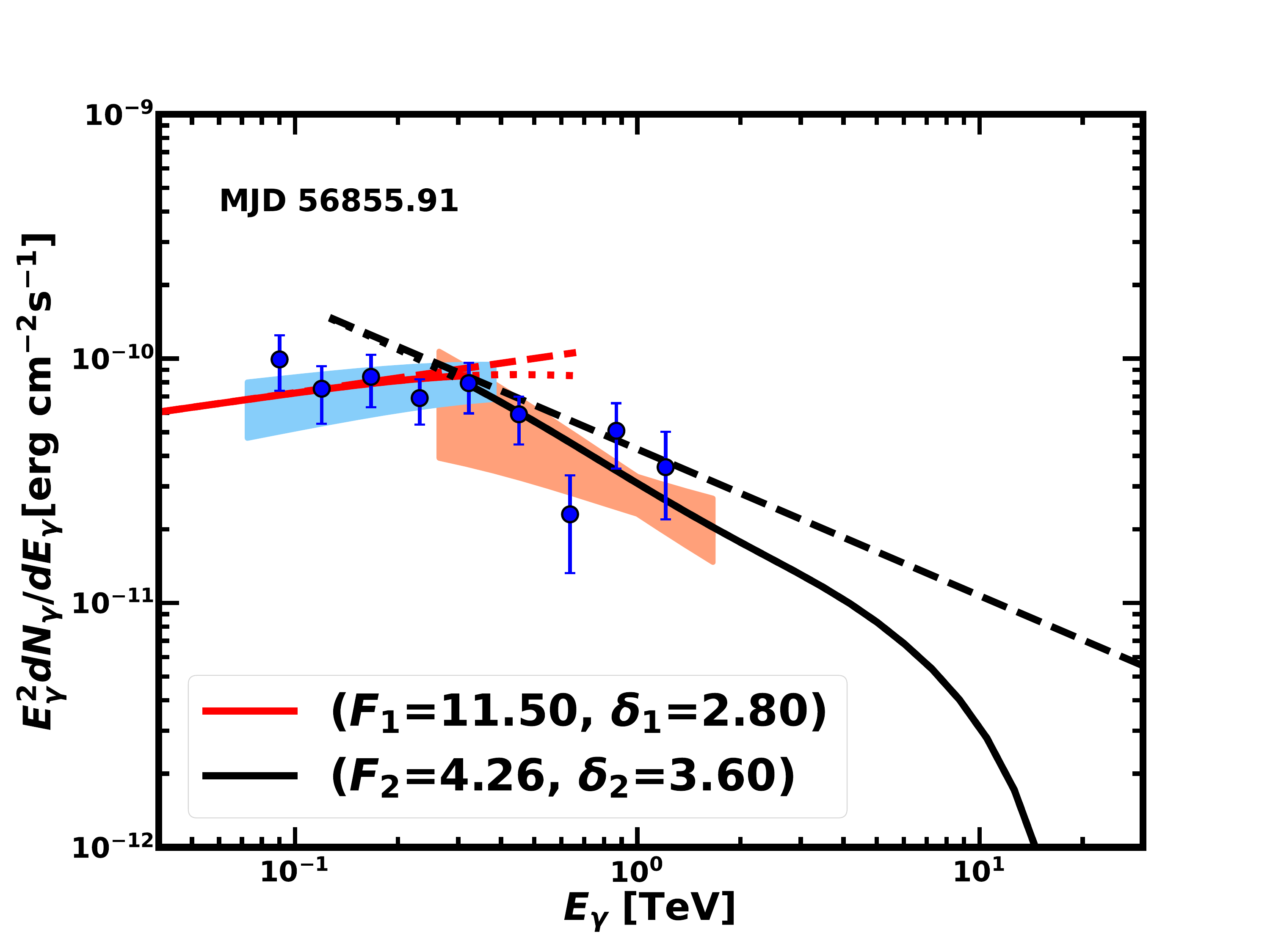}
        \label{Fig1b}
        
    \includegraphics[width=0.46\textwidth]{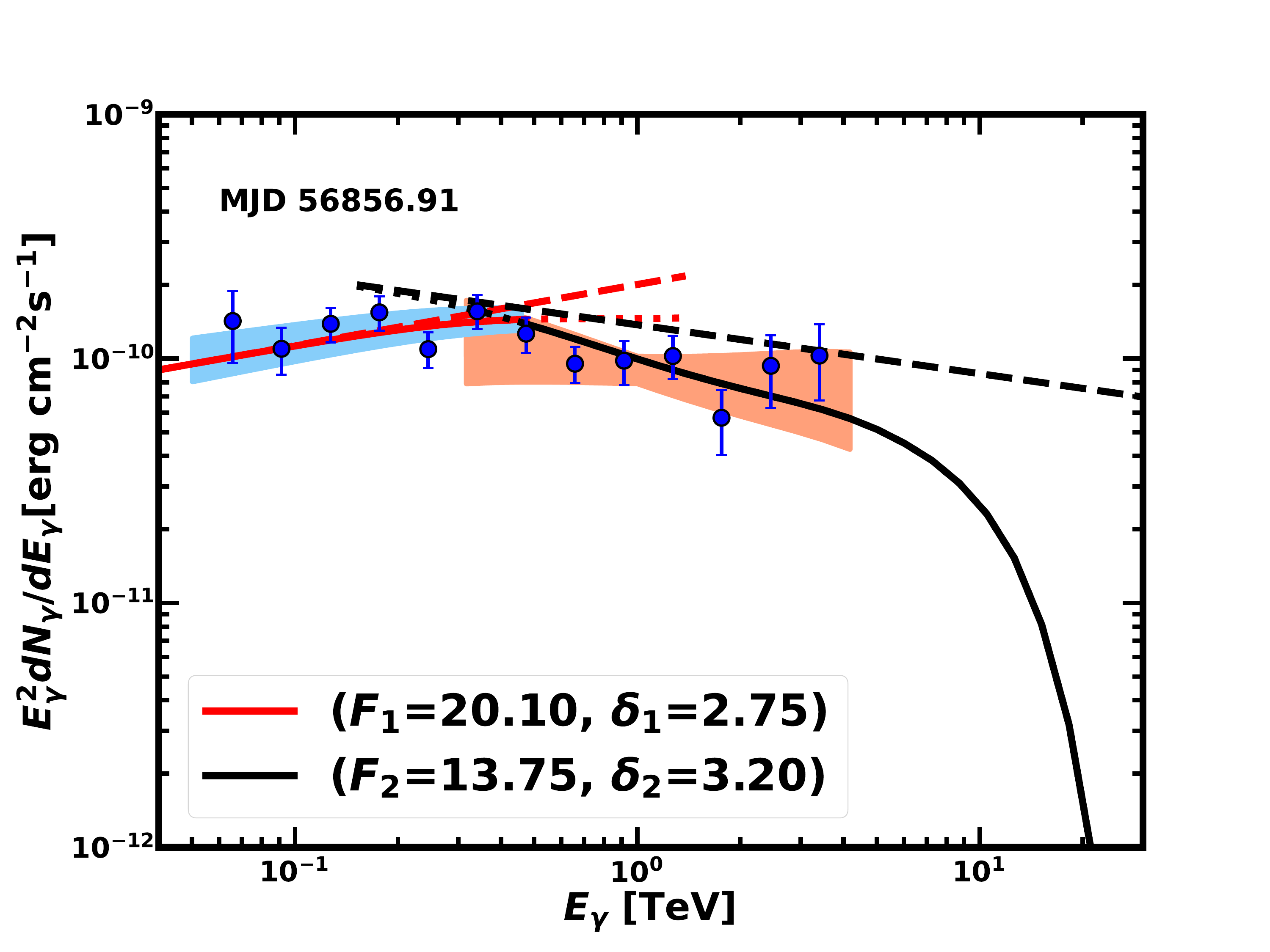}
        \label{Fig1c}
    \caption{The observed VHE spectra of Mrk 501 during the first three days are fitted with the two-zone photohadronic model. The normalization constants $F_1$ and $F_2$ are given in units of $10^{-11}\, \mathrm{erg\,cm^{-2}\,s^{-1}}$. The red and black solid curves represent the best fits for zone 1 and zone 2, respectively; dotted and dashed extension show their behaviors if each zone exist alone, The dashed red and black lines correspond to the intrinsic spectra of zone-1 and 2. The blue and salmon shaded regions indicate the $1\sigma$ confidence intervals for the best fits with ($F_1$, $\delta_1$) and ($F_2$, $\delta_2$), respectively. Figs.~\ref{fig:Fig2}, \ref{fig:Fig3}, \ref{fig:Fig4}, and \ref{fig:Fig5} follow the same definitions and units.}
    \label{fig:Fig1}
  \end{center}
\end{figure}

\begin{figure}
    \centering
    \includegraphics[width=1\linewidth]{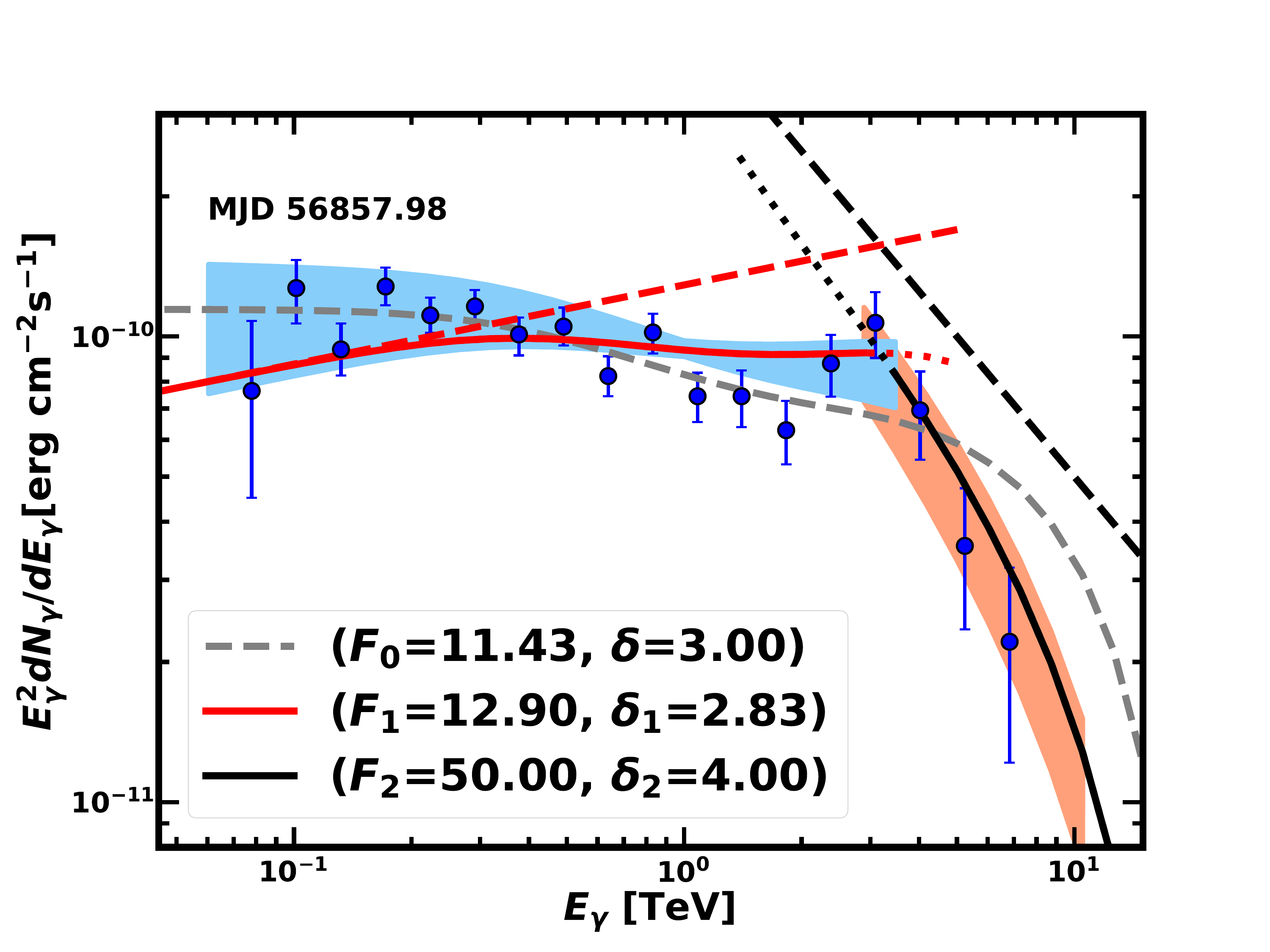}
    \caption{
    The VHE spectrum of Mrk 501 observed on MJD 56857.98. The red curve corresponds to zone-1 and the black curve to zone-2 (see Fig. \ref{fig:Fig1} for definitions). Unlike Fig. \ref{fig:Fig1}, this observation shows a notably sharp flux suppression above $\sim$ 3 TeV, indicating an abrupt cutoff in the high-energy component. The gray dotted curve is the standard photohadronic model (one-zone) fit.}
    \label{fig:Fig2}
\end{figure}

\begin{figure}
  \begin{center}
    \includegraphics[width=0.4\textwidth]{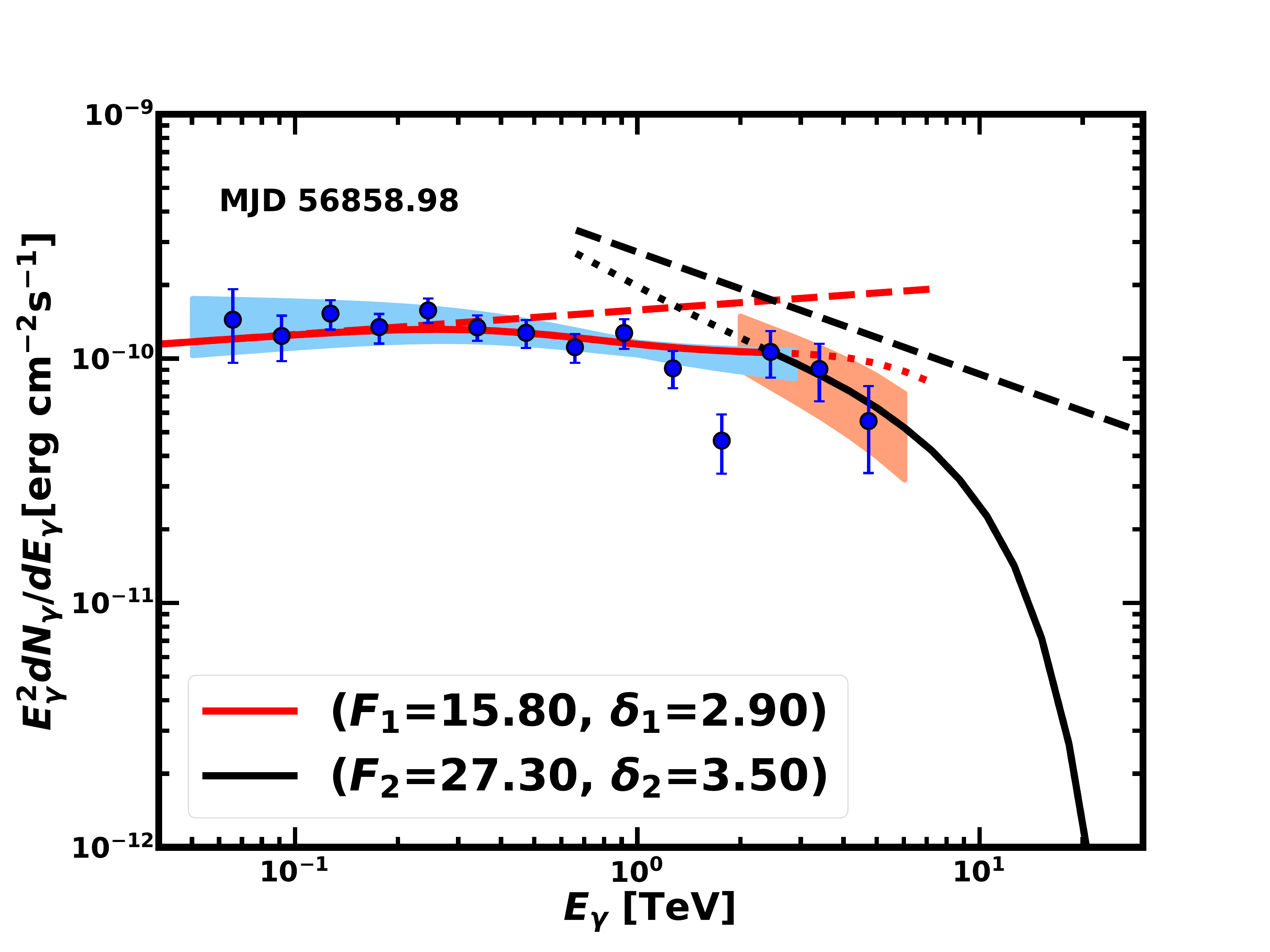}
        \label{Fig3a}
    \includegraphics[width=0.4\textwidth]{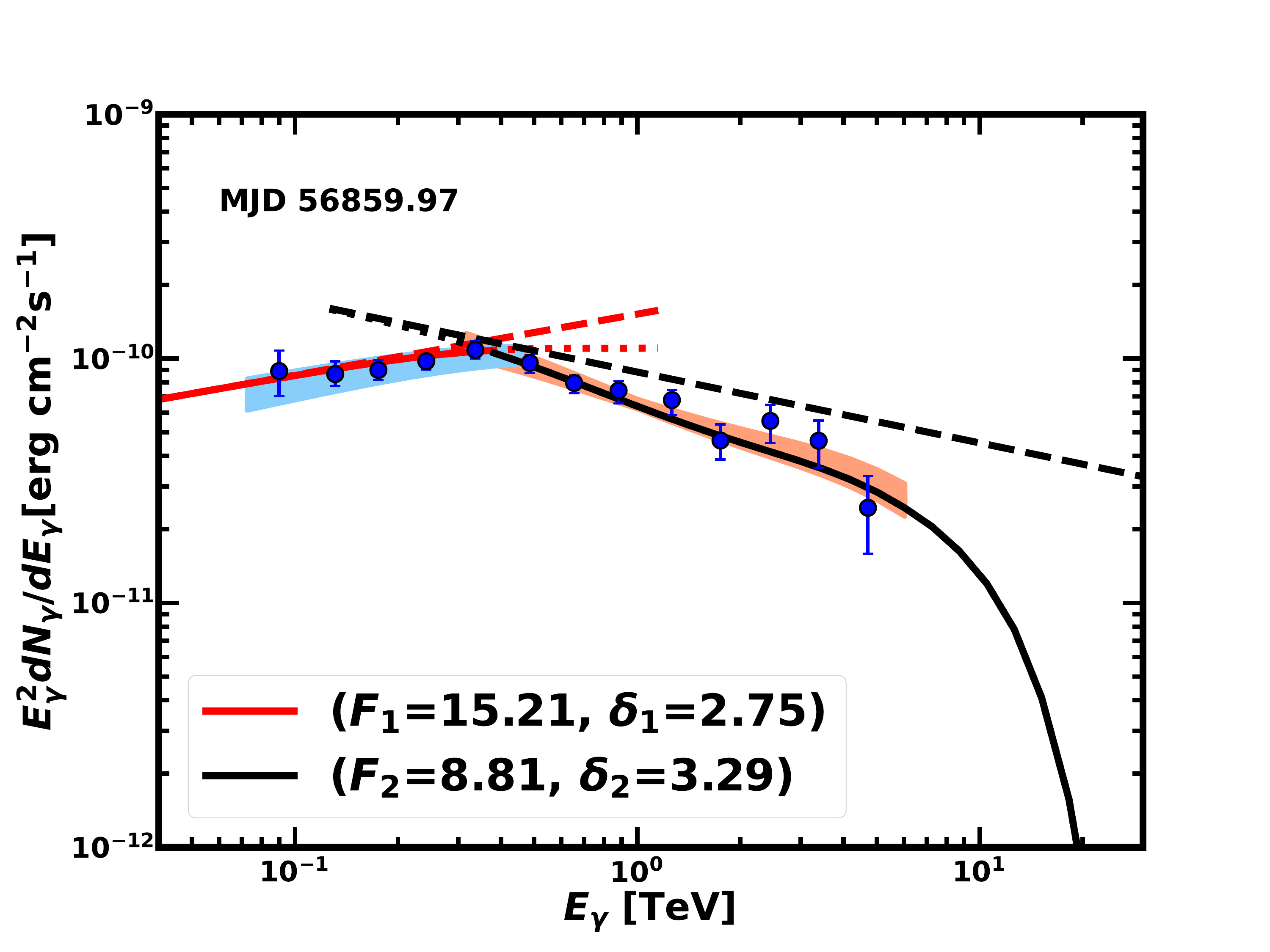}
        \label{Fig3b}
    \includegraphics[width=0.4\textwidth]{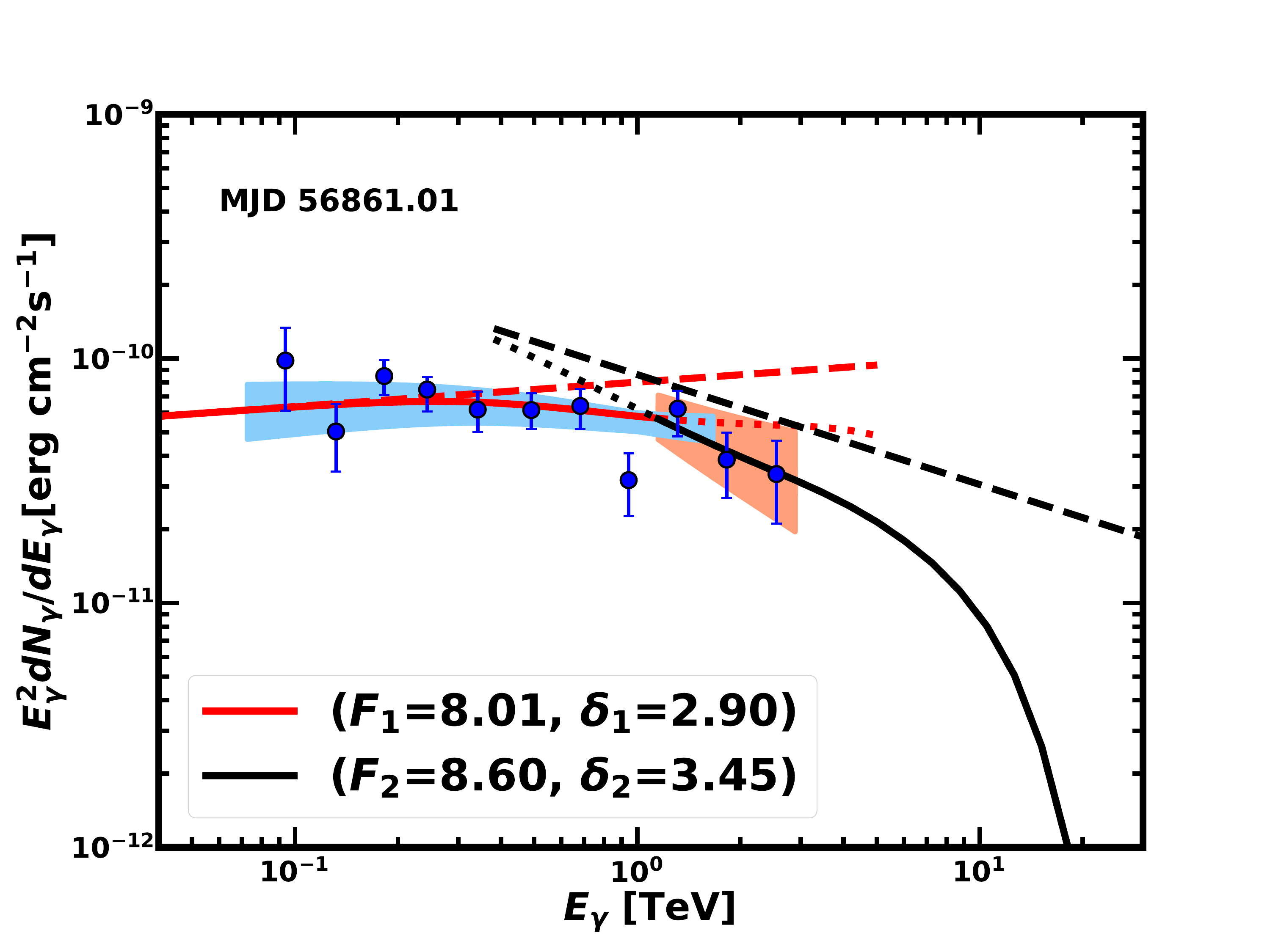}
        \label{Fig3c}
    \includegraphics[width=0.4\textwidth]{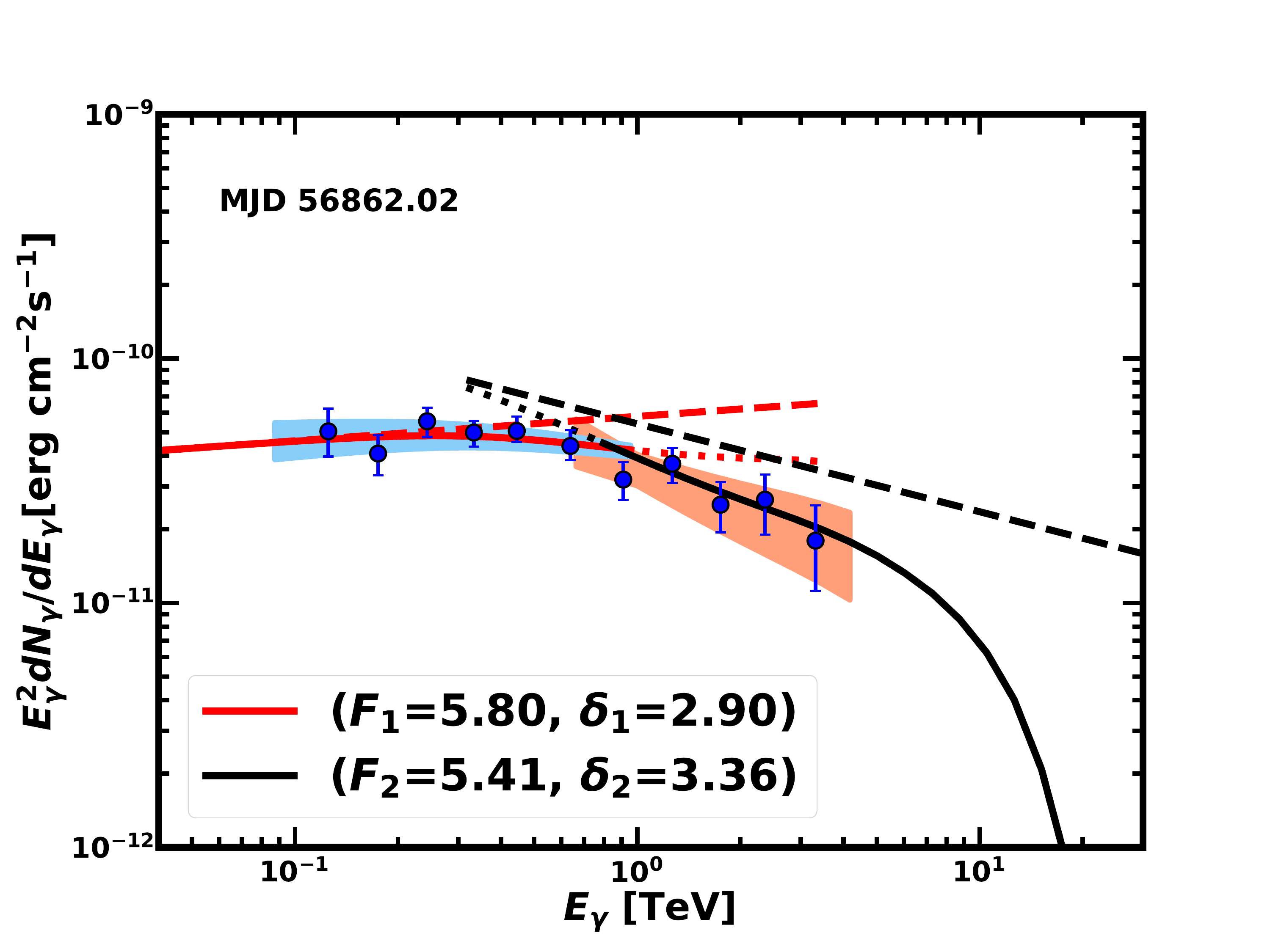}
        \label{Fig3d}
\caption{
    The observed VHE spectra on MJD 56858.98, 56859.97, 56861.01 and 56862.02 are fitted with the two-zone photohadronic model (see Fig. \ref{fig:Fig1} for definitions). These spectra exhibit strong day-to-day variations in the cutoff energy. Additionally, none of the subfigures display a smooth or flat spectral shape, indicating significant dynamical changes in the background photon field during this period.}
    \label{fig:Fig3}
  \end{center}
\end{figure}

\begin{figure}
  \begin{center}
  \includegraphics[width=0.4\textwidth]{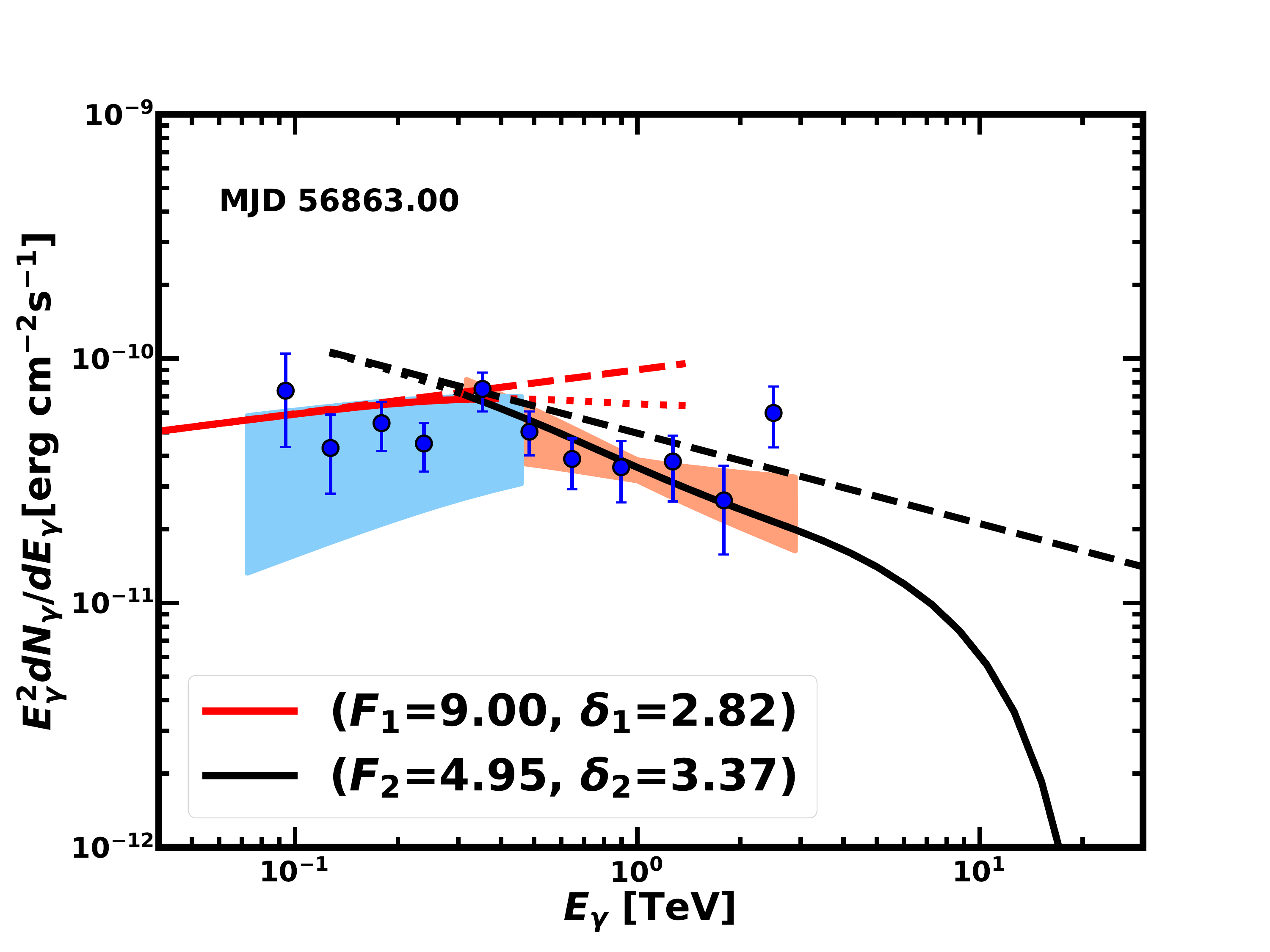}
        \label{Fig4a}
    \includegraphics[width=0.4\textwidth]{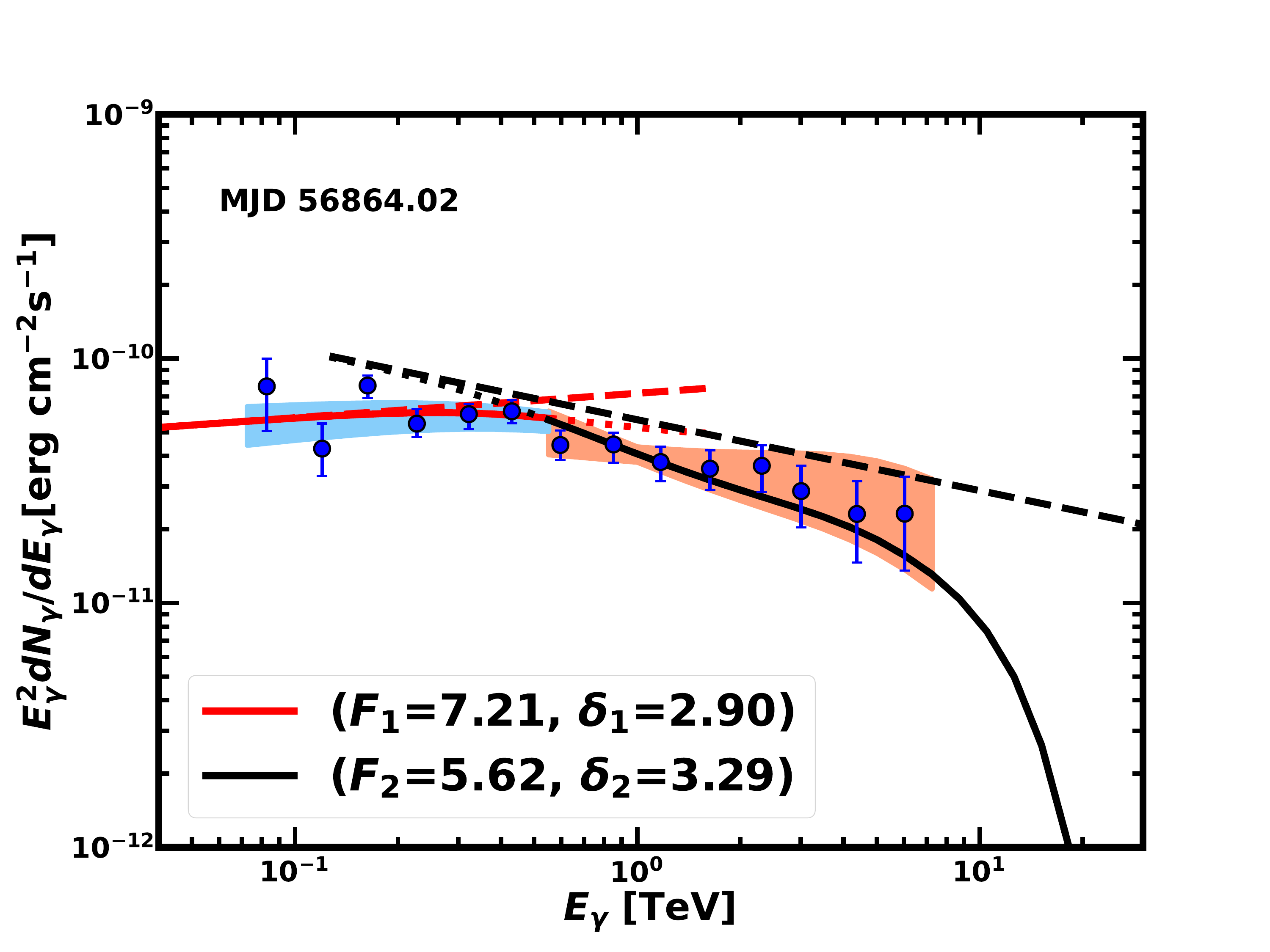}
        \label{Fig4b}
    \includegraphics[width=0.4\textwidth]{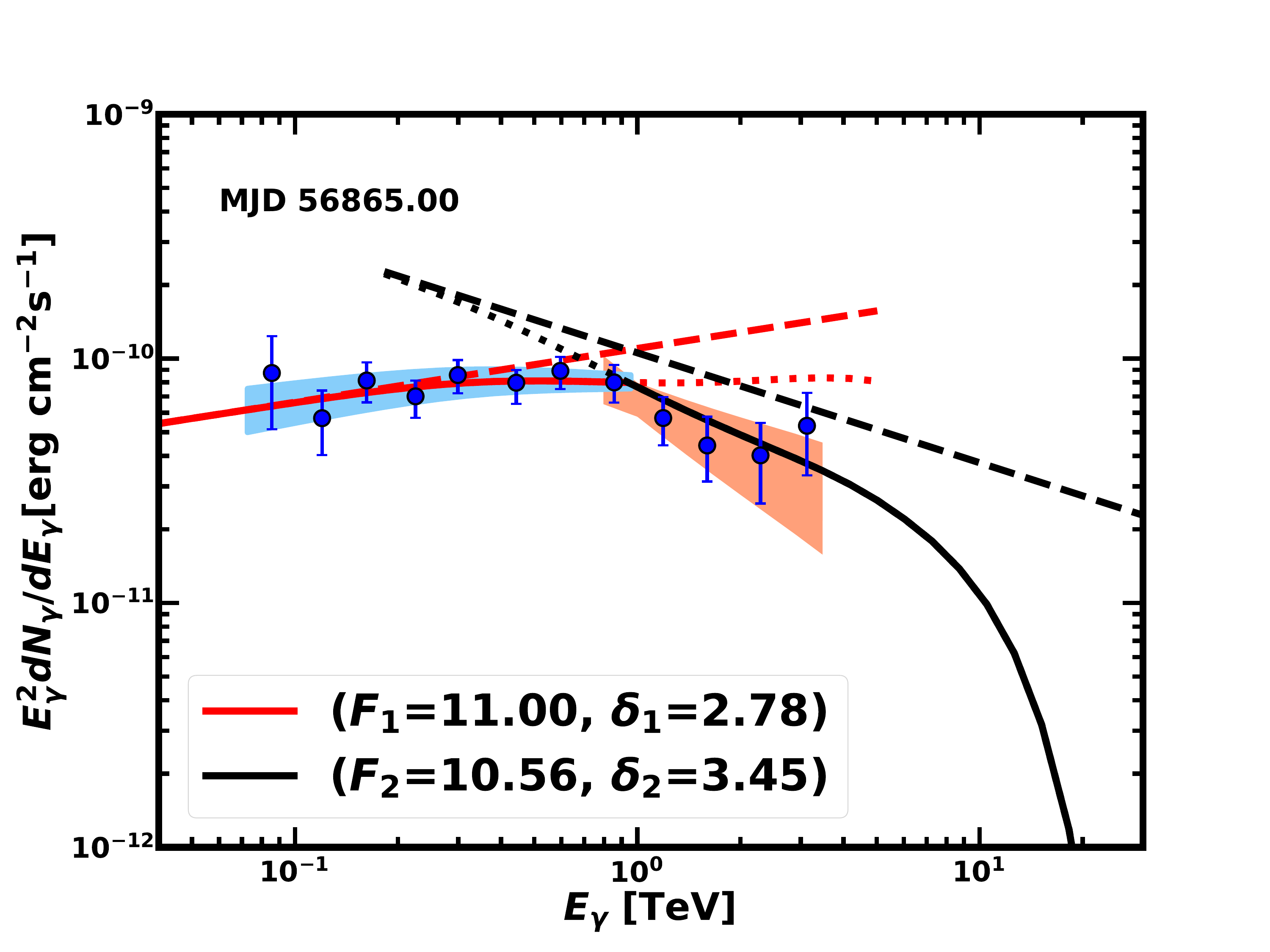}
        \label{Fig4c}
    \includegraphics[width=0.4\textwidth]{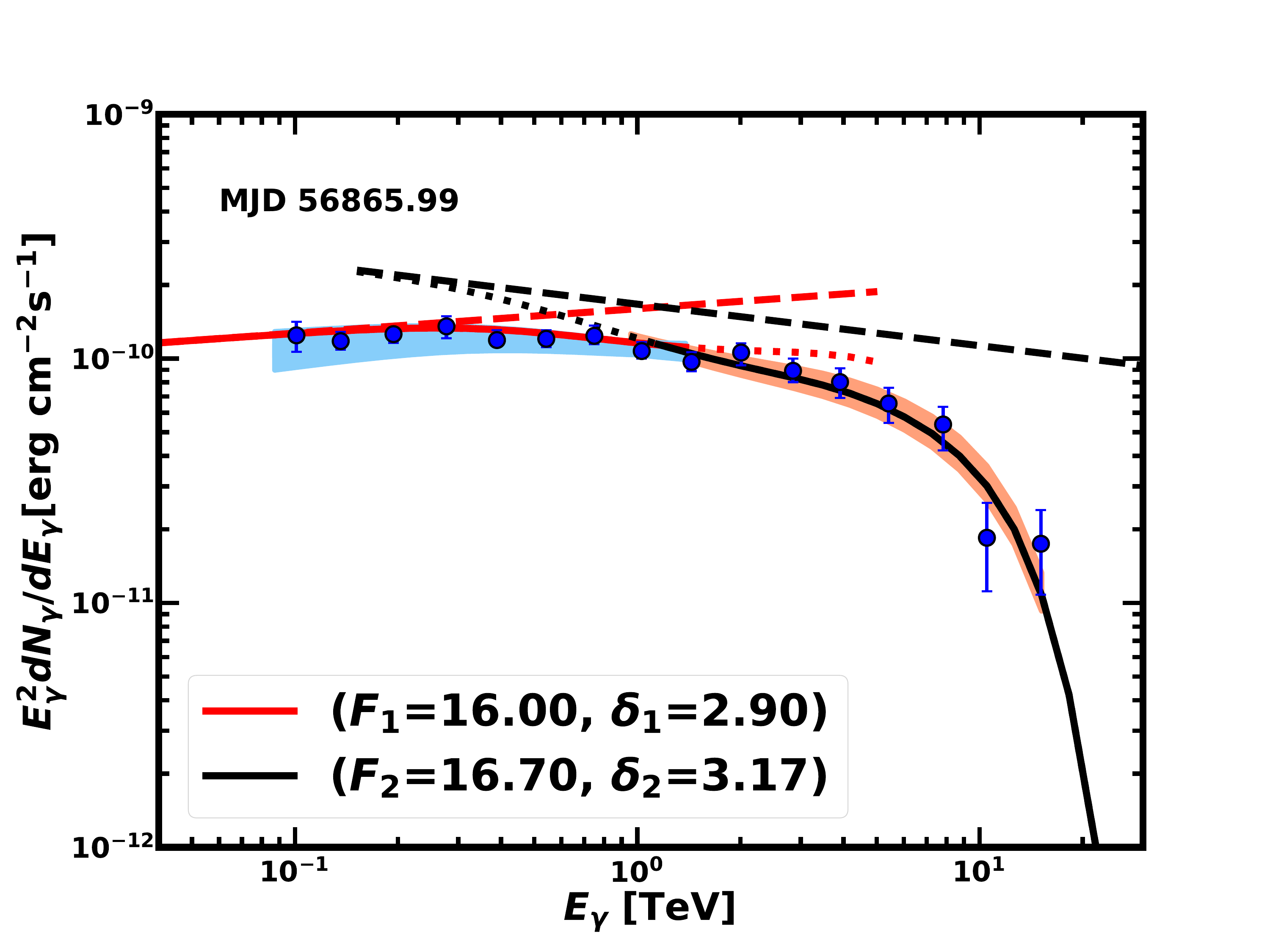}
        \label{Fig4d}
    \caption{The VHE spectra observed on MJD 56863.00, 56864.02, 56865.00 and 56865.99 are fitted with the two-zone photohadronic model (see Fig. \ref{fig:Fig1} for definitions). 
    For MJD 56863.00, the last data point near $\sim$ 2.5 TeV should not be interpreted as a spectral peak, as it lacks a definite spectral index for zone-2.}
    \label{fig:Fig4}
  \end{center}
\end{figure}

\begin{figure}
  \begin{center}

    \includegraphics[width=0.4\textwidth]{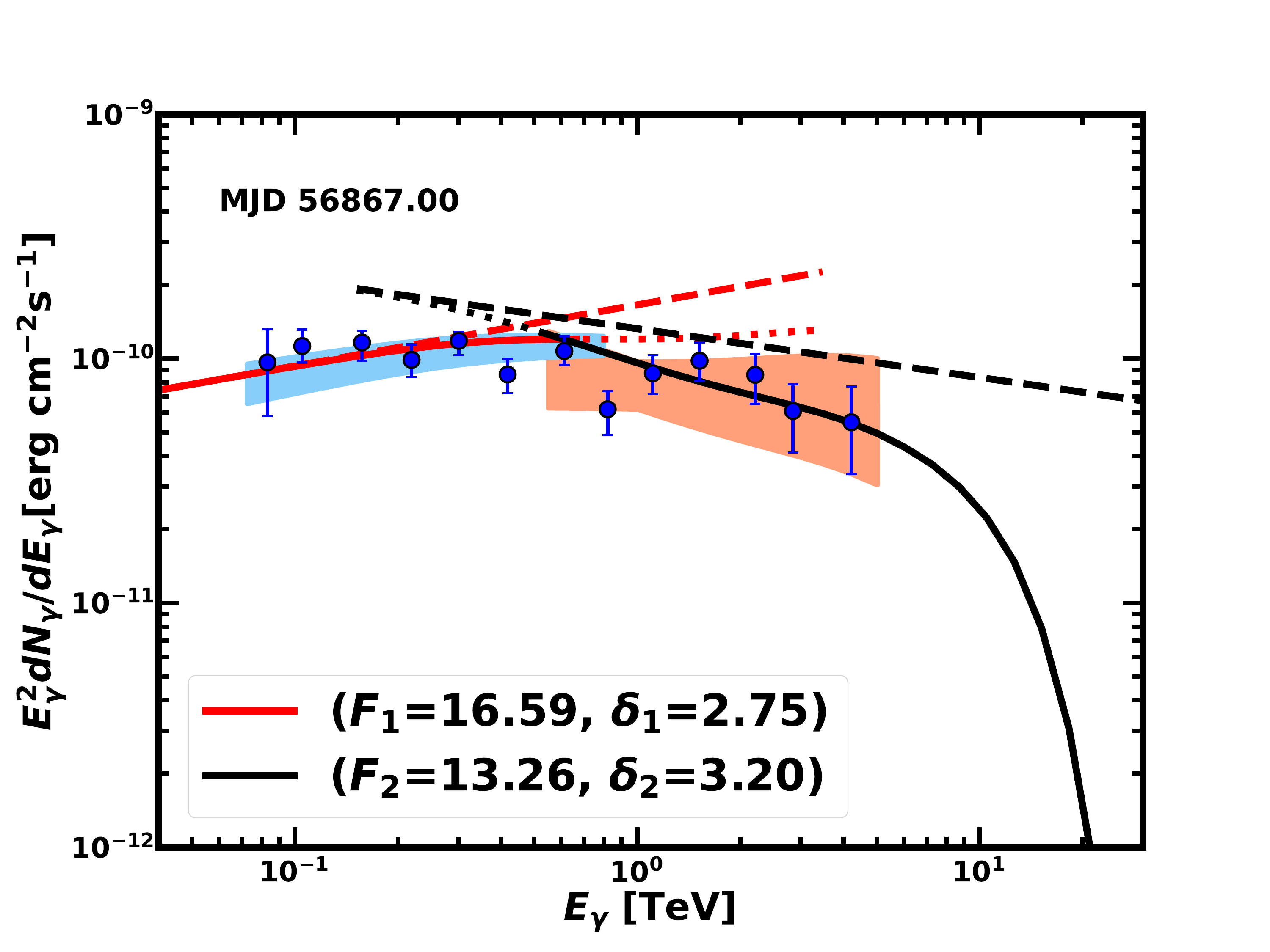}
        \label{Fig5a}
    \includegraphics[width=0.4\textwidth]{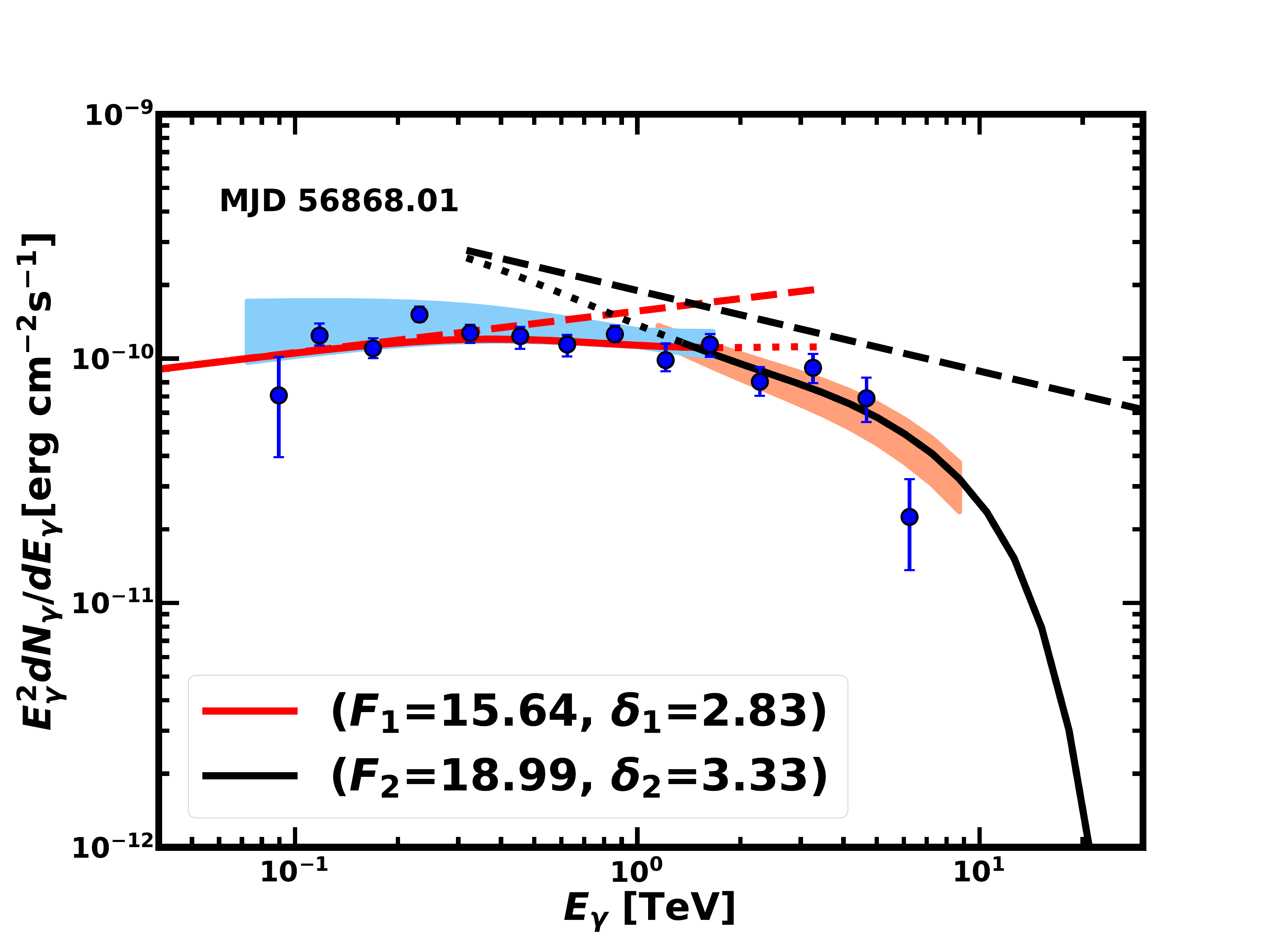}
        \label{Fig5b}
    \includegraphics[width=0.4\textwidth]{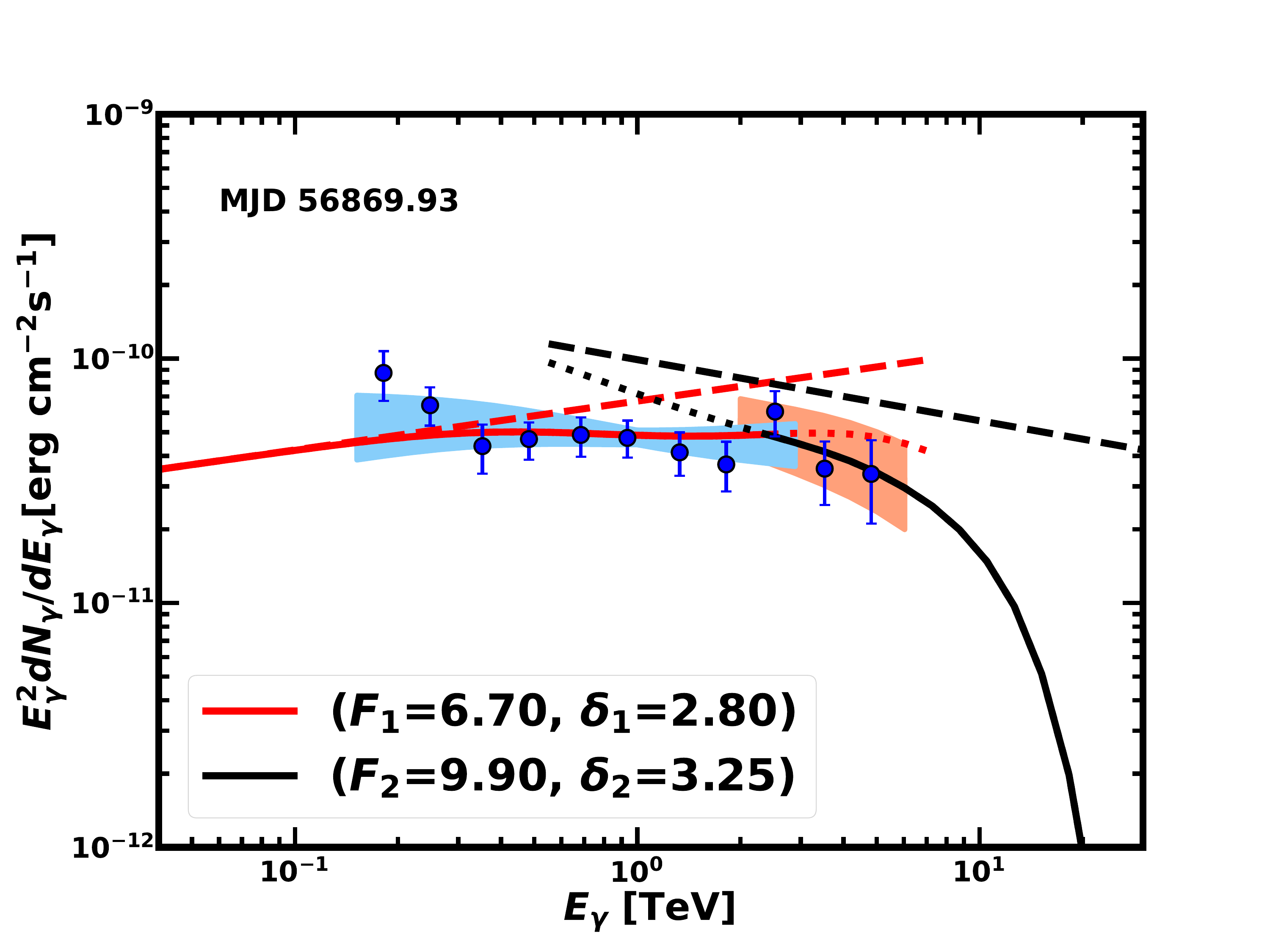}
    \label{Fig5c}
    \caption{The observed VHE spectra on MJD 56867.00, 56868.01, and 56869.93 are fitted with the two-zone photohadronic model (see Fig. \ref{fig:Fig1} for definitions). Compared to previous days, the spectra in this period appear smoother, with a gradual increase in the cutoff energy $E^c_{\gamma}$. On MJD 56869.93, zone-2 has only two data points above $E^c_{\gamma}=2.38$ TeV. However, the spectrum in zone-2 is well described within the existing two-zone framework of the photohadronic model.}
    \label{fig:Fig5}
  \end{center}
\end{figure}

\begin{figure}
{\centering
{\includegraphics[scale=0.172]{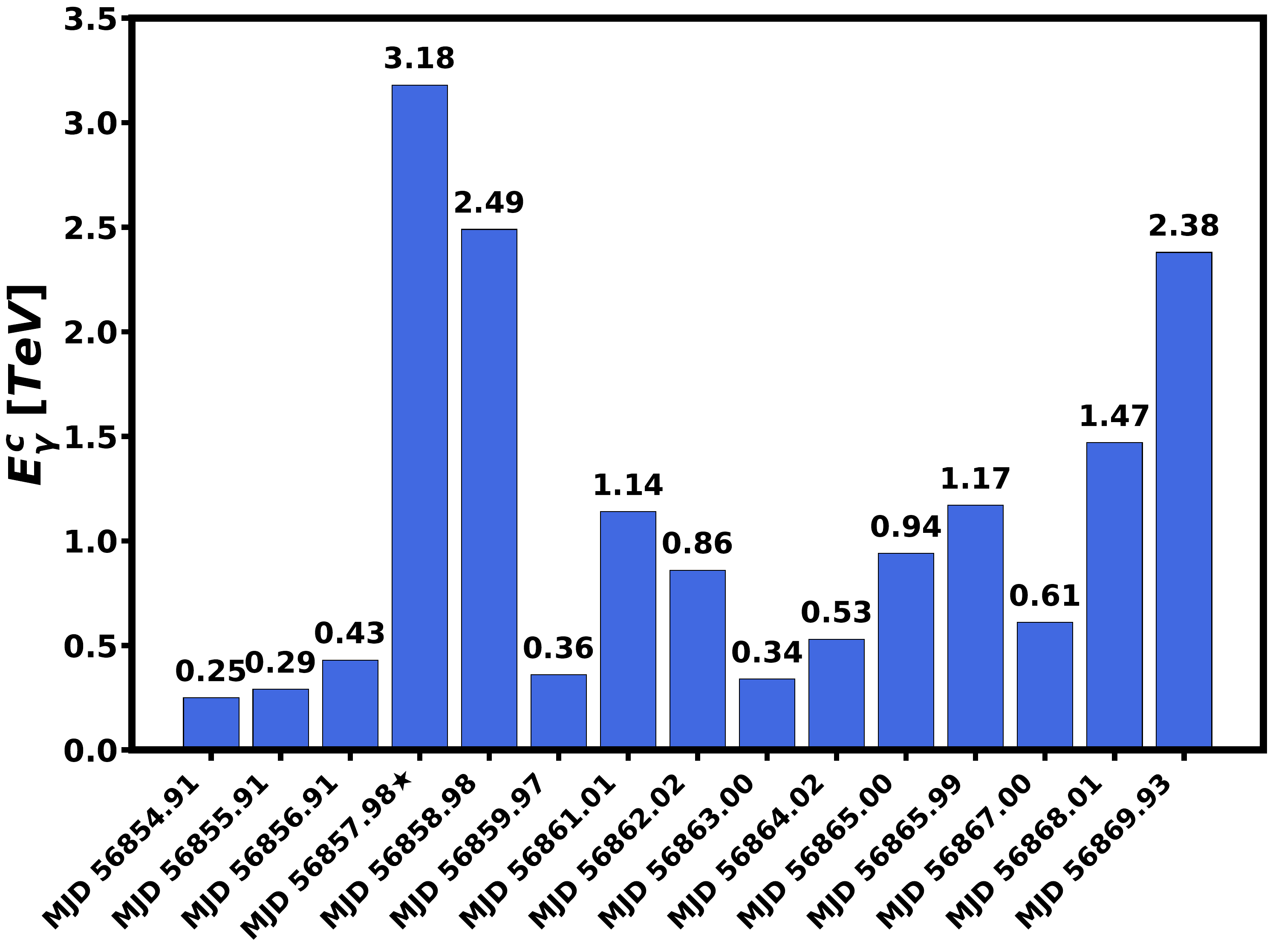}}
\par}
\caption{The cutoff energy $E^c_{\gamma}$ in the two-zone photohadronic model is shown for every flaring event. MJD 56857.98$^*$ corresponds to the day when a narrow peak-like feature at 3.18 TeV was observed and this is also shown in Fig.~\ref{fig:Fig7} (a) and Fig.~\ref{fig:Fig7} (b).}
\label{fig:Fig6}
\end{figure}

\begin{figure}
\centering
\includegraphics[scale=0.172]{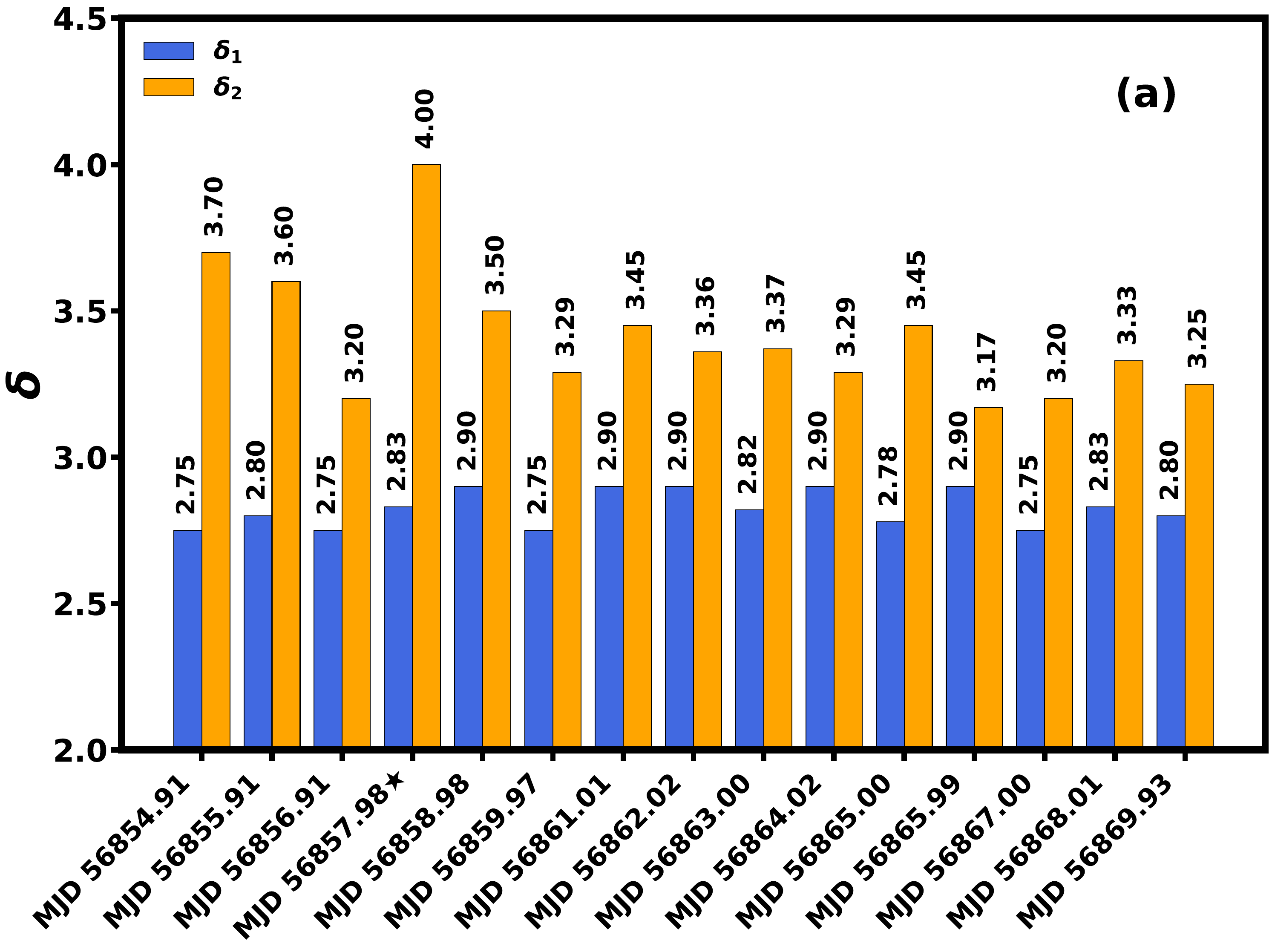}
\includegraphics[scale=0.172]{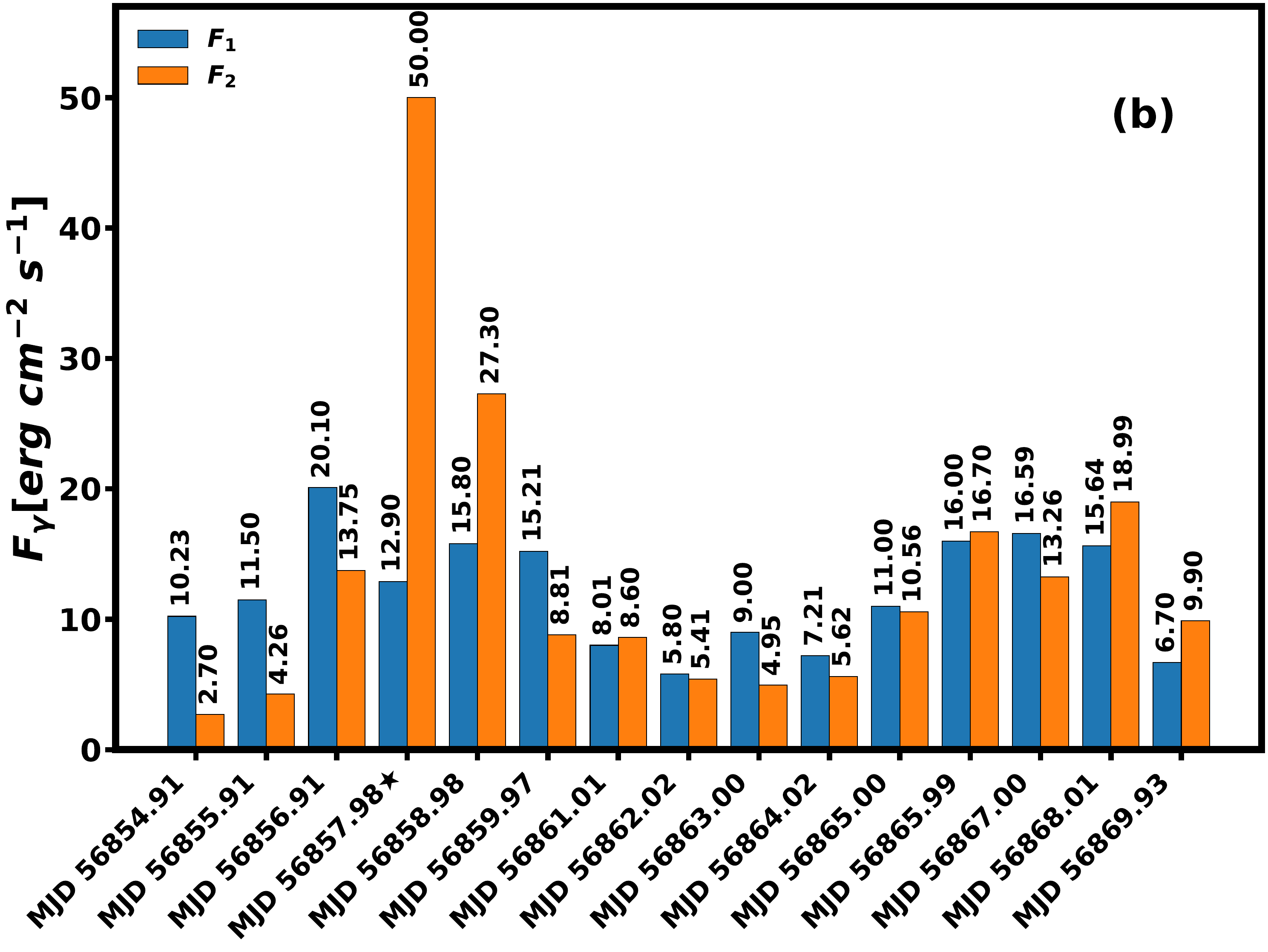}
\caption{(a) The best fit values of the spectral indices $\delta_1$ and $\delta_2$ for every flaring event are shown. (b) The best fit values of the normalization factors $F_1$ and $F_2$ in two-zone photohadronic model are shown for all the days.}
\label{fig:Fig7}
\end{figure}

\section{Results}

The Mrk 501 is an extensively studied HBL in multiple wavelengths since its discovery and its VHE spectra are well explained by leptonic and hadronic models \cite{2011ApJ...727..129A,2015ApJ...812...65F,2017A&A...603A..31A,2020A&A...637A..86M,2001APh....15..121M,2003APh....18..593M}. However, during its VHE flaring in the years 1997, 2005 and 2012 \cite{2020ApJ...901..132S}, it was observed that the synchrotron peak and the SSC peak were shifted towards higher energies. The synchrotron peaks were much above $10^{17}$ Hz, signaling that the emissions were tEHBL-like, thus, presented difficulties in explaining the VHE spectra using the leptonic model. Also, the standard photohadronic model was inadequate to explain these VHE SED.  We use two-zone photohadronic model to explain these spectra well. The flaring of Mrk 501 during July 2014 as shown in Table~\ref{table1}, had exactly the same behavior that of the flaring of 1997, 2005 and 2012, when the synchrotron peak was in the EHBL regime (above $10^{17}$ Hz) and the X-ray outburst was correlated with the VHE emission. So, we employ the two-zone photohadronic model to explain these 15 VHE spectra (from July 16 to July 31) on a day-by-day basis. 
A narrow peak-like feature at $\sim 3$ TeV was found for the first time on July 19, the day when the highest X-ray flux was also observed. This feature is inconsistent with the traditional interpretation of the VHE spectra using different analytic functions. We shall also discuss in detail this narrow peak-like feature.


\begin{table}
\centering
\caption{A summary of the parameters used for the calculation of the VHE spectra for all the days. The flux normalization factors $F_1$ and $F_2$ are in units of $10^{-11}\, \mathrm{erg\, cm^{-2}\, s^{-1}}$. The cutoff energy $E^c_{\gamma}$ is in TeV. The $1\sigma$ confidence level (CL) interval for the parameters $F_1$, $\delta_1$, $F_2$ and $\delta_2$ are shown.
}
\setlength{\tabcolsep}{3pt} 
\renewcommand{\arraystretch}{1.3}
\begin{tabular}{llllll}
\hline
 MJD & $F_1$ & $\delta_1$ & $F_2$ & $\delta_2$ & $E^c_{\gamma}$ \\
\hline 
56854.91 & 10.23$^{+0.9}_{-1.7}$ & 2.75$^{+0.05}_{-0.1}$ & 2.70$^{+0.2}_{-0.6}$ & 3.70$^{+0.08}_{-0.2}$ & 0.25 \\
56855.91 & 11.50$^{+0.7}_{-1.6}$ & 2.80$^{+0.05}_{-0.08}$ & 4.26$^{+0.3}_{-1.1}$ & 3.60$^{+0.08}_{-0.4}$ & 0.29 \\
56856.91 & 20.10$^{+1.4}_{-1.4}$ & 2.75$^{+0.07}_{-0.03}$ & 13.75$^{+0.4}_{-2.8}$ & 3.20$^{+0.05}_{-0.4}$ & 0.43 \\
56857.98$^{\star}$ & 12.90$^{+0.5}_{-0.3}$ & 2.83$^{+0.2}_{-0.003}$ & 50.00$^{+3.6}_{-8.4}$ & 4.00$^{+0.2}_{-0.04}$ & 3.18 \\
56858.98 & 15.80$^{+0.3}_{-1.5}$ & 2.90$^{+0.1}_{-0.01}$ & 27.30$^{+3.7}_{-4.6}$ & 3.50$^{+0.2}_{-0.1}$ & 2.49 \\
56859.97 & 15.21$^{+0.3}_{-1.3}$ & 2.75$^{+0.01}_{-0.06}$ & 8.81$^{+0.6}_{-0.2}$ & 3.29$^{+0.03}_{-0.09}$ & 0.36 \\
56861.01 & 8.01$^{+0.3}_{-1.1}$ & 2.90$^{+0.08}_{-0.04}$ & 8.60$^{+1.8}_{-1.2}$ & 3.45$^{+0.3}_{-0.3}$ & 1.14 \\
56862.02 & 5.80$^{+0.2}_{-0.4}$ & 2.90$^{+0.07}_{-0.04}$ & 5.41$^{+0.2}_{-1.3}$ & 3.36$^{+0.2}_{-0.2}$ & 0.86 \\
56863.00 & 9.00$^{+0.1}_{-3.7}$ & 2.82$^{+0.01}_{-0.4}$ & 4.95$^{+0.4}_{-0.6}$ & 3.37$^{+0.07}_{-0.4}$ & 0.34 \\
56864.02 & 7.21$^{+0.2}_{-0.7}$ & 2.90$^{+0.04}_{-0.05}$ & 5.62$^{+0.3}_{-0.4}$ & 3.29$^{+0.03}_{-0.4}$ & 0.53 \\
56865.00 & 11.00$^{+0.8}_{-0.6}$ & 2.78$^{+0.05}_{-0.05}$ & 10.56$^{+0.4}_{-2.7}$ & 3.45$^{+0.4}_{-0.2}$ & 0.94 \\
56865.99 & 16.00$^{+0.1}_{-1.7}$ & 2.90$^{+0.01}_{-0.09}$ & 16.70$^{+0.5}_{-1.09}$ & 3.17$^{+0.05}_{-0.11}$ & 1.17 \\
56867.00 & 16.59$^{+0.4}_{-2.2}$ & 2.75$^{+0.04}_{-0.05}$ & 13.26$^{+0.1}_{-4.7}$ & 3.20$^{+0.03}_{-0.4}$ & 0.61 \\
56868.01 & 15.64$^{+2.5}_{-0.1}$ & 2.83$^{+0.2}_{-0.01}$ & 18.99$^{+1.3}_{-1.3}$ & 3.33$^{+0.1}_{-0.04}$ & 1.47 \\
56869.93 & 6.70$^{+0.3}_{-0.6}$ & 2.80$^{+0.21}_{-0.04}$ & 9.90$^{+1.9}_{-1.3}$ & 3.25$^{+0.1}_{-0.1}$ & 2.38 \\
\hline 
\end{tabular}
\begin{minipage}{\columnwidth}
\vspace{0.1cm}
\small Notes: $^{\star}$ A peak-like feature observed around 3 TeV on MJD 56857.98.
\end{minipage}
\label{table1}
\end{table}
During these 15 consecutive days observation period, there were no strictly simultaneous observations in X-rays and gamma rays. However, 
there was no significant intra-night variability in the X-ray and the VHE gamma-ray bands. So, one can assume that the X-ray and the VHE gamma-ray observations were simultaneous.

As stated above, the flaring of Mrk 501 during July 2014 was tEHBL-like, so, we shall use the two-zone photohadronic model to fit the VHE gamma-spectra. The VHE spectrum on MJD 56854.91 was observed in the energy range $0.13\, \mathrm{TeV}\le E_{\gamma} \le 1.3\, \mathrm{TeV}$ and 
is fitted well with $\delta_1=2.75$, $F_1=10.23\times 10^{-11}\, \mathrm{erg\,cm^{-2}\,s^{-1}}$ for zone-1 and $\delta_2=3.70$, $F_2=2.70\times 10^{-11}\, \mathrm{erg\,cm^{-2}\,s^{-1}}$ for zone-2 as shown in Fig. \ref{fig:Fig1}. The cutoff energy between zone-1 and zone-2 is $E^c_{\gamma}=0.25$ TeV. For zone-1, $\delta_1=2.75$ corresponds to a high emission state and the intrinsic VHE photon flux $F_{\gamma,in}\propto E^{0.30}_{\gamma}$. But, for zone-2, $\delta_2=3.70$ and $F_{\gamma,in}\propto E^{-0.70}_{\gamma}$ which is consistent with the fast fall of the flux above $E^c_{\gamma}$. Also, our fits to both zones are very good.
Similarly, on MJD 56855.91, the best fit to the spectrum is obtained for
$\delta_1=2.80$, $F_1=11.50\times 10^{-11}\, \mathrm{erg\,cm^{-2}\,s^{-1}}$ for zone-1 and $\delta_2=3.60$, $F_2=4.26\times 10^{-11}\, \mathrm{erg\,cm^{-2}\,s^{-1}}$ for zone-2 as shown in Fig. \ref{fig:Fig1}. Here the cutoff energy $E^c_{\gamma}=0.29$ TeV, slightly higher than the previous day. On the third day (MJD 56856.91), the zone-1 is in high emission state with $\delta_1=2.75$ and in zone-2, the flux decreases slower as compared to the previous day with $\delta_2=3.20$ and $E^c_{\gamma}=0.43$ TeV (Fig. \ref{fig:Fig1}). It can be observed that $E^c_{\gamma}$ has increased slowly from 0.25 TeV, from the first day to 0.43 TeV, to the third day. For each day, the value of $E^c_{\gamma}$ is adjusted depending on the behavior of the spectrum. The good fit values of the parameters $F_1$, and $\delta_1$ corresponding to zone-1 for each spectrum, the confidence interval at $1\sigma$ is shown in the blue shaded region. Similarly, for zone-2, the confidence interval at $1\sigma$ is shown in the salmon shaded region for good fit values of the parameters $F_2$, and $\delta_2$ for each spectrum.

A noticeable change in the spectrum was observed on MJD 56857.98, on the fourth day, as shown in Fig. \ref{fig:Fig2}. The VHE spectrum was observed in the energy range $0.08\, \mathrm{TeV}\le E_{\gamma} \le 6.8\, \mathrm{TeV}$. 
A peak-like feature is observed $\sim 3$ TeV and the spectrum falls rapidly above this energy, which is a new feature of this spectrum. Like previous days, we have used the two-zone photohadronic model to fit the spectrum. The zone-1 is still in the high emission state with $\delta_1=2.83$, but the cutoff energy has drastically jumped to $E^c_{\gamma}=3.18$ TeV. In the energy range $0.5\, \mathrm{TeV}\lesssim E_{\gamma} \lesssim 3.18\, \mathrm{TeV}$ (zone-1), the EBL corrected spectrum is attenuated 
with an inflection point at $E_{\gamma}\sim 1.82$ TeV. 

In zone-2, the spectrum falls extremely fast, which can be fitted with $\delta_2=4.0$, as shown in Fig. \ref{fig:Fig2}. This is the first time we have observed such a rapid fall in the VHE spectrum in zone-2 for the tEHBL-like behavior of an HBL with a very large value of $\delta_2$. 
For $E^c_{\gamma} \,<\,3.18$ TeV the fitted VHE spectrum is in high emission state and increases slowly. However, for $E^c_{\gamma}\, > 3.18$ TeV, the spectrum falls faster, resulting in a mild peak-like feature but not very prominent as claimed by the MAGIC Collaboration.

For comparison, we also fit the spectrum with the standard photohadronic model and the best fit values of the parameters are $\delta=3.0$ and $F_0=11.43\times 10^{-11}\, \mathrm{erg\,cm^{-2}\,s^{-1}}$. This curve falls smoothly with no peak-like feature and in the high energy limit it falls slower than the observed spectrum (Fig. \ref{fig:Fig2}). Moreover, $\delta=3.0$ corresponds to low emission state and its intrinsic spectrum is flat which is common during a quiescent state of the blazar. This clearly shows that we can not use the standard photohadronic model (one-zone) to fit the spectrum of MJD 56857.98.

Previously, we had studied the tEHBL-like behavior of Mrk 421, Mrk 501, and 1ES 1959+650 in different epochs using the two-zone photohadronic model, where we could explain the spectra very well and the maximum value of $\delta_2$ was 3.5~\cite{2020ApJ...901..132S,2021ApJ...906...91S,2021ApJ...914..120S}. We had never observed such a fast fall in the spectrum in zone-2 and a cutoff energy as high. In this context, the VHE spectrum of Mrk 501 on MJD 56857.98 is unique and probably the first of its type. The peak-like feature around $E^c_\gamma$ is observed to be due to a slight increase in the flux in zone-1 and then a rapid drop in the spectrum for $E_\gamma > E^c_\gamma$. 
The blue shaded region to the left and the salmon shaded region to the right in Fig.~\ref{fig:Fig2} correspond to the 
$1\sigma$ confidence interval for the good fit values of the parameters in zone-1 ($F_1$ and $\delta_1$), and in zone-2 ($F_2$ and $\delta_2$) respectively.

The VHE spectra from July 20 to July 24 (MJD 56858.98 to MJD 56862.02) are fitted very well, which are shown in Fig. \ref{fig:Fig3} along with their 1$\sigma$ confidence intervals for zone-1 and zone-2.
It can be seen that, the spectral index $\delta_1$ in zone-1 is always in the range $2.6 < \delta_1 < 3.0$, corresponding to a high emission state. Also, in zone-2 we have $ 3.0 < \delta_2 \le 3.5$ as observed in previous analysis to the VHE spectra of tEHBL-like behavior of several sources, including Mrk 501. On MJD 56858.98, the $E^c_{\gamma}=2.49$ TeV is high but the spectrum in zone-2 does not fall rapidly. So, we do not see any peak-like structure. On July 21 (MJD 56859.97), the cutoff energy drops to very low, $E^c_{\gamma}=0.36$ TeV.

From July 25 to July 28 (MJD 56863.00 to MJD 56865.99), the spectra are fitted extremely well and are shown in Fig. \ref{fig:Fig4} with their respective parameters $F_1$, $F_2$, $\delta_1$ and $\delta_2$. We have also shown the 1$\sigma$ confidence intervals for zone-1 and zone-2 for all.
Throughout the observation period, the maximum VHE photon energy observed is on MJD 56865.99, which is $E_{\gamma}=15.1$ TeV. 
In Fig. \ref{fig:Fig5}, the VHE spectra of MJD 56867.00 to MJD 56869.93 (July 29 to July 31) with their 1$\sigma$ confidence intervals to zone-1 and zone-2 are shown. Here also, the spectra are fitted very well. On July 31, again the cutoff energy $E^c_{\gamma}=2.38$ TeV, which is high. However, the spectrum in zone-2 falls smoothly.

We have shown the best fit values of $E^c_{\gamma}$ for all the 15 days observations in Fig. \ref{fig:Fig6}. It can be seen that for the flaring on MJD 56857.98, the $E^c_{\gamma}$ is the maximum and on next day it decreases slightly and then a free fall on MJD 56859.97 to $E^c_{\gamma}=0.36$ TeV. Also, we observe a oscillating behavior in the $E^c_{\gamma}$ values, which can be seen in Fig. \ref{fig:Fig6}. Similarly, we have shown the best fit values of the VHE photon spectral indices $\delta_1$ and $\delta_2$ for zone-1 and zone-2 respectively in Fig. \ref{fig:Fig7}(a). As the spectra in zone-1 are in high emission states, the values of  $\delta_1$ are very similar for the whole period. However, in zone-2, we observe a variation in $\delta_2$ and again, the largest value of $\delta_2=4.0$ is on MJD 56857.98.

In Fig. \ref{fig:Fig7}(b), the best fit values of the normalization factors $F_1$ and $F_2$ are plotted for the whole observation period. It can be seen that the variation in $F_1$ is minimal, however, there is a large variation in $F_2$ and again, on MJD 56857.98, the variation is the highest. These analyses clearly show that the VHE spectrum of MJD 56857.98 is very distinct from the rest. The best fit values of the parameters in the two-zone photohadronic model for all the 15 days observations are summarized in Table \ref{table1}.

On MJD 56857.98, the best fit to zone-2 is achieved for $\delta_2=4.0$. 
With such a large $\delta_2$, the spectrum above $E^c_{\gamma}=3.18$ TeV falls rapidly. On this day, the highest energy photon observed is $E_{\gamma}=6.8$ TeV. 
This photon will be produced from the interaction of 68 TeV proton ($10\, E_{\gamma}$) with the SSC seed photon in the low energy tail region. Assuming that this seed photon has minimum energy $\epsilon_{\gamma} \simeq 7.5\times 10^{20}$ Hz ($\sim 3.1$ MeV) ~\cite{2020A&A...637A..86M}, the minimum bulk Lorentz factor 
can be estimated from Eq.(\ref
{eq:kinproton}) which is found to be 
$\Gamma\simeq 26.5$ and this is consistent with the expectation. Again, using this $\Gamma$ and the cutoff energy $E^c_{\gamma}=3.18$ TeV in Eq.(\ref{eq:kinproton}), the corresponding cutoff energy for the SSC seed photon is $\epsilon^c_{\gamma}\sim 6.6$ MeV.
By adopting the proton spectral index $\alpha=2.0$~\cite{1993ApJ...416..458D}, we get $\beta_2=2.0$ and $\beta_1=0.83$. The low energy tail region of the SSC flux (zone-2) behaves as $\Phi_{SSC}\propto
\epsilon^{2}_{\gamma}$ in the energy range $3.1\, \mathrm{MeV}\lesssim \epsilon_{\gamma} \lesssim 6.6\, \mathrm{MeV}$ and for $\epsilon_{\gamma} \gtrsim 6.6\, \mathrm{MeV}$ (zone-1), $\Phi_{SSC}\propto
\epsilon^{0.83}_{\gamma}$. 
The increase of $\Phi_{SSC}$ is quadratic as a function of the seed photon energy $\epsilon_{\gamma}$ in zone-2 and implies that the intrinsic VHE flux falls as 
$F_{\gamma,in}\propto E^{-2.0-2.0+3}_{\gamma} \propto E^{-1}_{\gamma}$. Similarly, in zone-1, the intrinsic flux increases mildly as  $F_{\gamma,in} \propto E^{-2.0-0.83+3}_{\gamma} \propto E^{0.17}_{\gamma}$.


The cutoff energy $E_{\gamma}^c$ divides the VHE spectrum into zone-1 and zone-2 with different photon spectral indices. Thus, for a given spectrum, the position of $E_{\gamma}^c$ in the two-zone photohadronic model is decided depending on the spectral behavior. If the spectrum shows a peak-like structure as in the case of MJD 56854.91 or MJD 56855.91 then the fitting is straightforward. However, in some cases the spectrum seems to be smooth and in these situations it is tricky to find the cutoff energy as in the case of MJD 56865.99. On MJD 56854.91 we do not have data above 1.3 TeV but a distinct peak at $\sim 0.25$ TeV and the two-zone photohadronic model fits well to the spectrum. Thus, the change in the spectral behavior is important to decide the position of the cutoff energy. Of course, availability of more data points in the high energy regime will help to improve the fit in zone-2, but not necessarily to fix the value of $E_{\gamma}^c$.

\subsection{Comparison with flaring events of 2005 and 2012}

The spectral analysis of the VHE flaring events during July 16 to July 31, 2014 (total 15 days) shows that the flaring events in zone-1 were in high emission states like a standard HBL spectrum which we have studied previously. Also, the spectra in zone-1 are very similar to the flaring events of Mrk 501 observed during 2005 May-July and 2012 June~\cite{2020ApJ...901..132S}. Similar spectra were also observed from Mrk 421 in 2010 March~\cite{2021ApJ...914..120S}. However, we observed a large variation in the VHE spectra of Mrk 501 in zone-2 during 2014 flaring event, particularly on July 19, 2014 \cite{2020A&A...637A..86M}. On this day, the spectral index $\delta_2=4.0$ was observed to be very large, corresponding to a sharp fall in the VHE gamma-ray spectrum above a large cutoff energy $E^c_{\gamma}=3.18$ TeV. 

During previous flaring events of Mrk 501 in 2005 and 2012, we did observe a variation in the spectra in zone-2, $3.1 \lesssim \delta_2 \lesssim 3.5$. However, this time the variation was observed to be abrupt. For example, on MJD 56856.91, the $E^c_{\gamma}=0.43$ TeV and $\delta_2=3.2$ and on next day MJD 56857.98 it jumped to 3.18 TeV and 4.0 respectively. Such large values were not seen before in any of the tEHBL flaring events. On this day, the intrinsic flux in zone-1 increased as $E^{0.17}_{\gamma}$ and decreased as $E^{-1}_{\gamma}$ in zone-2. Thus, at $E^c_{\gamma}=3.18$ TeV there is a peak-like structure in the intrinsic spectrum but manifests as a milder peak in the observed spectrum due to EBL smoothening effect.
As the spectral indices $\beta_1$ and $\beta_2$ are the distribution of the background seed photons in zone-1 and zone-2 respectively, the tEHBL-like behavior of Mrk 501 must be attributed to SSC band where $\beta_2=2.0$ (zone-2). 
We compare the values of $F_2$, $\delta_2$ and $E^c_{\gamma}$ of MJD 56857.98 with other days as summarized in Table \ref{table1} and plotted in Figs. \ref{fig:Fig6} and \ref{fig:Fig7}. This comparison clearly shows that
the spectrum of MJD 56857.98 is different from the rest.

Previously, it was shown that the leptonic modeling of many HBL SEDs have deep valleys at the junction of synchrotron spectrum and the SSC spectrum with large $\beta$ value \cite{2020ApJ...901..132S,2021ApJ...906...91S,2021ApJ...914..120S,2017EPJC...77...18S,2018EPJC...78..557S}. It is possible that during these extreme flaring events, the inverse Compton scattering of high energy electrons with the self-produced background synchrotron photons produce SSC photons in the narrow energy range of $3.1\, \mathrm{MeV}\lesssim \epsilon_{\gamma} \lesssim 6.6\, \mathrm{MeV}$ (zone-2), which are piled up and the flux behaves as $\Phi_{SSC}\propto \epsilon^2_{\gamma}$
and for $\epsilon_{\gamma} \gtrsim 6.6\, \mathrm{MeV}$ (zone-1) the flux is flatter behaving as $\Phi_{SSC}\propto \epsilon^{0.83}_{\gamma}$. Thus, the high energy protons interacting with these background SSC photons will produce the observed VHE gamma-ray spectrum.

\section{Discussion}

Mrk 501 is a HBL which is extensively studied in all wavelengths. Both leptonic models and hadronic models explain its spectra well. But the VHE emission during 2005, 2012 and 2014, when the synchrotron peak had shifted to frequencies above $10^{17}$ Hz, the VHE spectra were difficult to explain. Actually, during these flaring periods, Mrk 501 behaved like tEHBL and its spectra were complex. Similar tEHBL-like behavior also observed from Mrk 421 and 1ES 1959+650. In previous studies, we used the two-zone photohadronic model and explained very well the tEHBL-like behavior of these sources. In the present work we have used the two-zone photohadronic model again to explain the tEHBL-like behavior of Mrk 501 for the 15 days during July 16 to 31, 2014. It is observed that the spectra in zone-1 were like standard HBL and were in high emission states implying that the spectral indices in this region were in the range $2.6 < \delta_1 < 3.0$. During the VHE flaring period of 2005 and 2012 in Mrk 501, this same behavior was observed. However, the spectral behavior in zone-2 of the flaring in 2014 is much more complex and different than in the previous periods. Of particular importance is the VHE flaring event on MJD 56857.98 (July 19, 2014), when the cutoff energy $E^c_{\gamma}=3.18$ TeV and the spectral index in zone-2 was $\delta_2=4.0$. This high value of $\delta_2$ implies that the intrinsic spectrum falls as $F_{\gamma,in}\propto E^{-1}_{\gamma}$ for  $E_{\gamma} > E^c_{\gamma}$. In zone-1, $F_{\gamma,in}\propto E^{0.17}_{\gamma}$, a slowly increasing flux up to $E^c_{\gamma}$. In the photohadronic model we predict a mild peak-like structure at $E^c_{\gamma}=3.18$ TeV in the observed spectrum. On this day, the maximum observed photon energy was $E_{\gamma}=6.8$ TeV, corresponding to a minimum bulk Lorentz factor $\Gamma\simeq 26.5$ which is consistent with the expectation from an EHBL.
Except on 19th July, such spectral behavior was not observed on any other days. Also, such spectral behavior was not observed in the past from any of the HBLs, including Mrk 501.
It is expected that there may also be narrow TeV feature in the VHE spectrum on July 18 (one day before) and/or on July 20 (one day after). However, the photohadronic model does not predict any such feature on these days.

In the context of the two-zone photohadronic model, the VHE flaring event of MJD 56857.98 is due to the interaction of high energy protons with the low energy SSC background photons. In zone-1, protons of energy $E_p\lesssim 31.8\,\mathrm{TeV}$ interact with the SSC photons of energy $\epsilon_{\gamma} \gtrsim \,6.6\,\mathrm{MeV}$ to produce the observed VHE gamma-ray spectrum of energy $E_{\gamma} \lesssim 3.18\, \mathrm{TeV}$. In zone-2, protons in the energy range $31.8\, \mathrm{TeV}\lesssim E_p \lesssim 68\, \mathrm{TeV}$ interact with the SSC photons in the energy range $3.1\, \mathrm{MeV}\lesssim \epsilon_{\gamma} \lesssim 6.6\, \mathrm{MeV}$ to produce the VHE gamma-ray spectrum in the energy range
$3.18\, \mathrm{TeV}\lesssim E_{\gamma} \lesssim 6.8\, \mathrm{TeV}$. The VHE spectra of other days are similar to the flaring event of MJD 56857.98 but have different spectral behavior of the SSC photons in zone-1 and zone-2 and also have different cutoff energy $E^c_{\gamma}$.

On MJD 56865.99 and MJD 56868.01, the spectra are relatively smooth which could probably be fitted with a single power-law. However, during the whole observation period, the synchrotron peak of Mrk 501 was always above $10^{17}$ Hz, a clear signature of transition to an EHBL. Thus, the VHE spectra must be dealt with two-zone photohadronic model~\cite{2020ApJ...901..132S,2021ApJ...906...91S,2021ApJ...914..120S}. On the contrary, by using a single power-law, there will be no consistent way of dealing with these spectra. The two-zone approach is therefore not an arbitrary choice, but a necessary condition to reproduce the spectral shape 
and their temporal evolution during different flaring periods when the synchrotron peak is above $10^{17}$ Hz.


It is to be noted that our two-zone photohadronic model explains well all the 15 spectra of the VHE flaring events of Mrk 501. Other nearby HBLs, which have already shown the tEHBL-like behavior in the past, may reincarnate again with vigor and behave like the flaring of July 19, 2014 of Mrk 501 or even more complex. However, many more flaring events from nearby HBLs have to be studied to establish this claim. With the present and/or future generation of IACTs it will be possible to monitor such flaring events in great detail to better understand their emission mechanisms and enrich theoretical modeling.  
   
\begin{acknowledgements}
The work of S.S. is partially supported by DGAPA-UNAM (Mexico) Project No. IN105326. A.U.P.O., D.I.P.S and G. S.-C. would like to thank SECIHTI (Mexico) for partial support. Partial support from CSU-Long Beach is gratefully acknowledged. Pedro Fernández de Córdoba's work is partially supported by the Spanish government through the RD projects PID2021-128676OB-I00 and PID2022-142407NB-I00, as well as by the Excellence Research Groups Program of the Generalitat Valenciana, Spain (PROMETEO CIPROM / 2023 / 32). 
\end{acknowledgements}


\bibliographystyle{spphys}

\bibliography{mrk501}

\end{document}